\documentclass[aps,floatfix,preprint,preprintnumbers,nofootinbib,superscriptaddress,natbib]{revtex4}
\usepackage{graphicx,float}% Include figure files
\usepackage{epsfig}		
\usepackage{dcolumn}% Align table columns on decimal point
\usepackage{bm}% bold math

\usepackage[all]{xy}
\usepackage{amsmath,upgreek}
\usepackage{amssymb}

\usepackage{pdfpages}
\usepackage{color}
\usepackage[colorlinks=true,
 linkcolor=red,
 urlcolor=purple,
 citecolor=blue,hyperindex]{hyperref}

\def\0{\mbox{\tiny $0$}}
\def\1{\mbox{\tiny $1$}}
\def\2{\mbox{\tiny $2$}}
\def\3{\mbox{\tiny $3$}}
\def\4{\mbox{\tiny $4$}}
\def\5{\mbox{\tiny $5$}}
\def\6{\mbox{\tiny $6$}}
\def\7{\mbox{\tiny $7$}}
\def\8{\mbox{\tiny $8$}}
\def\9{\mbox{\tiny $9$}}

\def\f14{\mbox{\tiny $\frac{1}{4}$}}

\def\bb#1{\mbox{\footnotesize $(#1)$}}

\def\bb#1{\mbox{\footnotesize $(#1)$}}
\def\mt#1{\mbox{\textsl{#1}}}
\def\mbf#1{\mbox{\boldmath$#1$}}
\def\bb#1{\mbox{\small $(#1)$}}

\DeclareMathOperator{\Tr}{\mbox{Tr}}

\renewcommand{\baselinestretch}{1.5}

\begin{document}

\title{Algebraic solutions for $SU(2)\otimes SU(2)$ Hamiltonian eigensystems: generic statistical ensembles and a mesoscopic system application}

\author{Alex E. Bernardini}
\email{alexeb@ufscar.br}
\affiliation{Departamento de F\'{\i}sica, Universidade Federal de S\~ao Carlos, PO Box 676, 13565-905, S\~ao Carlos, SP, Brasil}
\author{R. da Rocha}
\affiliation{Federal University of ABC, Center of Mathematics, Computing and Cognition, 09210-580, Santo Andr\'e, Brazil}\email{roldao.rocha@ufabc.edu.br}

\begin{abstract}
Solutions of generic $SU(2)\otimes SU(2)$ Hamiltonian eigensystems are obtained through systematic manipulations of quartic polynomial equations. An {\em ansatz} for constructing separable and entangled eigenstate basis, depending on the quartic equation coefficients, is proposed. Besides the quantum concurrence for pure entangled states, the associated thermodynamic statistical ensembles, their partition function, quantum purity and quantum concurrence are shown to be straightforwardly obtained. Results are specialized to a $SU(2)\otimes SU(2)$ structure emulated by lattice-layer degrees of freedom of the Bernal stacked graphene, in a context that can be extended to several mesoscopic scale systems for which the onset from $SU(2)\otimes SU(2)$ Hamiltonians has been assumed.
\end{abstract}

\date{\today}
\maketitle

\section{Introduction}

Hamiltonian dynamics supported by a $SU(2) \otimes SU(2)$ group structure driven by two internal degrees of freedom have been considered in a broad set of physical systems, from relativistic frameworks for particle dynamics \cite{AlexC,n011,n011A,n011B,PRAPRA18,EPJC,Universe} to mesoscopic quantum information platforms \cite{n001,n002,n003,n004,n005,Nat06,graph015,graph01,graph02,graph03,graph04}. This encompasses, for instance, the algebra representation of Dirac bi-spinors for computing {\em spin and parity} intrinsic entanglement profiles encoded by such internal degrees of freedom (DoF) of a single particle \cite{PRAPRA18,n008,n010,n009,AoP2018}, the construction of gaussian localized Dirac cat states in phase space \cite{EPJP,PRA2021,PRA2023,PRA2024}, the investigation of ionic state states with $SU(2) \otimes SU(2)$ entanglement emulated by a complete set of quantum observables associated with their internal DoF \cite{n001,MeuPRA,New}, and also the $SU(2) \otimes SU(2)$ {\em lattice-layer} entanglement structure of Bernal stacked bilayer graphene \cite{Predin,MeuPRB,PRBPRB18} obtained for the quantum system described by a tight-binding Hamiltonian.

Considering such an ample scenario, a generalized understanding of the universal correlational structure of $SU(2)\otimes SU(2)$ Hamiltonian eigensystems is expected. From the perspective of the Lie subalgebra $SU(2)\otimes SU(2) \subset SL(2,\mathbb{C})\otimes SL(2,\mathbb{C})$, representations of $sl(2,\mathbb{C})\oplus sl(2,\mathbb{C})$, the Lie algebra of the $SL(2,\mathbb{C})\otimes SL(2,\mathbb{C})$ group, are irreducible\footnote{Their representations correspond to tensor products between linear complex representations of $sl(2,\mathbb{C})$.}. Unitary irreducible representations of $SU(2)\otimes SU(2)$ are then tensor products of unitary representations of $SU(2)$, which exhibits a one-to-one correspondence with the group $SL(2,\mathbb{C})\otimes SL(2,\mathbb{C})$ and, considering a simply connected group, the same correspondence with the algebra $sl(2,\mathbb{C})\oplus sl(2,\mathbb{C})$.
As a consequence, {\em inequivalent representations} of $SU(2) \otimes SU(2)$ follows straightforwardly, and provides to $SU(2) \otimes SU(2)$ Hamiltonian structures such a universal property (see the Appendix I) for encompassing several theoretical features, as those identified in the above mentioned physical systems.

In this work, the onset of solutions of generic $SU(2)\otimes SU(2)$ Hamiltonian eigensystems obtained through systematic manipulations of quartic polynomial equations suggests an {\em ansatz} for constructing separable and entangled orthonormal eigenstate basis for the associated Hamiltonian. As a natural extension, the associated thermodynamic ensemble partition function, and its corresponding quantum purity can be analytically obtained. As an application, results are specialized to a $SU(2)\otimes SU(2)$ structure emulated by lattice-layer degrees of freedom of the Bernal stacked graphene, in a context that can be extended to several mesoscopic scale systems for which the onset from $SU(2)\otimes SU(2)$ Hamiltonians have been assumed.

In particular, our results shall be specialized to the Bernal stacked graphene system, for which the associated Hamiltonian eigensystem solution is algebraically obtained, so as to extend and encompass the classes of solutions obtained in some previous works \cite{graph04,Predin,MeuPRB,PRBPRB18}, by classifying their entanglement pattern. 
In such a context, it is worth mentioning that graphene physics \cite{graphteo,graph05,graph06,graph07,graph08,graph010,graph011,graph012,graph013,graph014,deGracia:2023dfo}
has indeed enlarged the condensate matter horizon for the investigation of quantum entanglement, inserting additional relations with ground physics, which provides, for instance, a better understanding of the quantum Hall effect \cite{CondMatter01,CondMatter02,CondMatter03,CondMatter04,CondMatter05}, as well as straightforward applications to quantum computing processes \cite{QuantumInfoGraph01,QuantumInfoGraph02,QuantumInfoGraph03,QuantumInfoGraph04}. Bilayer graphene excitations, in its most stable configuration -- the AB (or Bernal) stacking \cite{graph03,graph04} -- can be indeed described in terms of the above discussed $SU(2)\otimes SU(2)$ Hamiltonian platform. More interestingly, if the tight binding Hamiltonian including both bias voltage and mass terms \cite{graph04,Predin} is written in the reciprocal space \cite{MeuPRB,PRBPRB18}, intrinsic entanglement properties arise, and therefore the Bernal stacked graphene quantum mechanical features can be properly interpreted in terms of the above addressed $SU(2)\otimes SU(2)$ symmetry parameters.

Following such a prescription, in Section II, a systematic procedure for obtaining the state associated spectra of $SU(2)\otimes SU(2)$ Hamiltonians is provided. A secular quartic equation depending on the driving parameters of Hamiltonian $SU(2)\otimes SU(2)$ interactions is obtained and two classes of eigensystem solutions are algebraically identified.
From such results, generalized expressions for the quantum concurrence of Hamiltonian eigenstates and for the quantized partition function of thermodynamic ensembles, including the generalized expressions related to quantum purity and quantum concurrence, are yielded. In Section III, results are specialized to the $SU(2)\otimes SU(2)$ structure emulated by lattice-layer degrees of freedom of the Bernal stacked graphene, for which quantum purity and quantum concurrence between the $SU(2)$ associated Hilbert space variables are computed. 
Finally, our conclusions are drawn in Section IV, and they address to the additional mesoscopic scale systems for which any onset $SU(2)\otimes SU(2)$ Hamiltonian can be investigated.

\section{$SU(2)\otimes SU(2)$ Hamiltonian eigensystems}

Let one consider the onset Hamiltonian, $H$, as composed by Hermitian $SU(2)$ invariant (two fold) tensor product contributions given by
\begin{equation}
\label{FirstH}
\hat{H} = \upsilon\,\hat{I}_{_{{ 2}}}^{(1)} \otimes\hat{I}_{_{{ 2}}}^{(2)} 
+ \sum_{i=1}^3 \alpha_i\, \hat{\sigma}_i^{(1)} \otimes\hat{I}_{_{{ 2}}}^{(2)} 
+ \sum_{j=1}^3 \beta_j \,\hat{I}_{_{{ 2}}}^{(1)}\otimes \hat{\sigma}_j^{(2)}
+ \sum_{i,j=1}^3 \omega_{ij}\, \hat{\sigma}_i^{(1)} \otimes \hat{\sigma}_j^{(2)},
\end{equation}
where $\upsilon$ is identified by
\begin{equation}
\label{AFirstH3}
\upsilon = \frac{1}{4}\Tr[\hat{H}],
\end{equation}
$\alpha_i,\, \beta_j$, and $\omega_{ij}$ are arbitrary coefficients, hats ``$~\hat{}~$'' denote operators, $\hat{\sigma}^{(1(2))}_i$ are the Pauli matrices, with $i,j = 1,\,2,\,3$ and 
\begin{equation}
\label{Pauli}
\hat{\sigma}^{(1(2))}_i \hat{\sigma}^{(1(2))}_j = \hat{I}_{_{{ 2}}}\,\delta_{ij}+\epsilon_{ijk}\,\hat{\sigma}^{(1(2))}_k,
\end{equation}
with $\hat{I}_{_{{ N}}}$ representing an $N$-dimensional identity matrix, $\delta_{ij}$ the Kronecker delta, and $\epsilon_{ijk}$ the Levi-Civita tensor. 
Assuming that the four-dimensional composite Hilbert space over which the $H$ eigenvectors span a basis is identified by $\mathbb{H} = \mathbb{H}^{(1)} \otimes \mathbb{H}^{(2)}$ with $\mbox{dim}\, \mathbb{H}^{(1)} = \mbox{dim}\, \mathbb{H}^{(2)} = 2$ one has\footnote{That is the group $Spin(2)$, which corresponds to a Clifford algebra of three real ($\mathbf{R}$) dimensions (one quaternionic ($\mathbf{H}$) dimension), since $SU(2)$ is the double covering group of $SO(3)$. The same is noticed for $SU(2)\otimes SU(2)$, $Spin(4)$, which is the double covering group of $SO(4)$, a Clifford algebra of six $\mathbf{R}$ dimensions (two $\mathbf{H}$ dimensions).} 
\begin{equation}
\label{FirstHcc}
\hat{H} \equiv \sum_{a,b} H^{(1)}_a\otimes H^{(2)}_b,
\end{equation}
which fits the $SU(2)\otimes SU(2)$ symmetry of the spanned basis (cf. Eq.~\eqref{FirstHcc} and $\mathbb{H} = \mathbb{H}^{(1)} \otimes \mathbb{H}^{(2)}$)\footnote{Concomitantly to the $SU(2)$ symmetry properties of each decoupled contribution, $H^{(1)}_a$ and $H^{(2)}_b$.}. 

Given the above preliminary conditions, one can try to algebraically solve the associated Hamiltonian eigensystem, 
\begin{equation}
\label{FirstH2}
\hat{H} \,\rho_{mn} = \varepsilon_{mn} \hat{I}_{_{{ 4}}}^{(1\otimes 2)} \, \rho_{mn},
\end{equation}
for state vectors denoted by $\rho_{mn}$ and eigenvalues $\varepsilon_{mn}$, with $m,\,n = 1,\,2$.
The strategy consists in obtaining quartic powers of $\hat{H}$. One can commence by identifying the traceless operator, 
\begin{equation}
\label{FirstH3}
\tilde{H} = \hat{H} - \upsilon\, \hat{I}_{_{{ 4}}},
\end{equation}
where the (Hilbert space) upper indices have been suppressed from this point.
The quadratic form of $\tilde{H}$ yields
\begin{eqnarray}
\label{SecondH}
\tilde{H}^2 &=& \alpha_i \alpha_j \,\hat{\sigma}_i\hat{\sigma}_j \otimes\hat{I}_{_{{ 2}}} 
+\beta_i \beta_j \,\hat{I}_{_{{ 2}}}\otimes \hat{\sigma}_i\hat{\sigma}_j
+ 2 \alpha_i \beta_j \, \hat{\sigma}_i \otimes \hat{\sigma}_j \nonumber\\
&&\qquad\qquad\qquad+\alpha_k \omega_{ij}\, \{\hat{\sigma}_k,\hat{\sigma}_i\} \otimes \hat{\sigma}_j
+\beta_k \omega_{ij} \,\hat{\sigma}_i\otimes\{\hat{\sigma}_k,\hat{\sigma}_j\}
+\omega_{ij}\omega_{kl} \,\hat{\sigma}_i\hat{\sigma}_k \otimes\hat{\sigma}_j\hat{\sigma}_l,
\end{eqnarray}
where the summations (from $1$ to $3$) have been denoted by repeated indices.
From Eq.~\eqref{Pauli}, and observing that $\{\hat{\sigma}_r,\hat{\sigma}_s\}=\hat{\sigma}_r\hat{\sigma}_s +\hat{\sigma}_s\hat{\sigma}_r = 2\,\hat{I}_{_{{ 2}}}\,\delta_{ij}$, Eq.~\eqref{SecondH} can be re-written as
\begin{eqnarray}
\label{SecondH2}
\tilde{H}^2 &=& \left\{\mbox{\boldmath$\alpha$}^2 + \mbox{\boldmath$\beta$}^2 + \Tr[\mbox{\boldmath$\omega$}\cdot\mbox{\boldmath$\omega$}^T] \right\} \hat{I}_{_{{ 2}}} \otimes\hat{I}_{_{{ 2}}} 
+\mathcal{A}_i\, \hat{\sigma}_i \otimes\hat{I}_{_{{ 2}}} 
+\mathcal{B}_j\,\hat{I}_{_{{ 2}}} \otimes\hat{\sigma}_j
+\mathcal{W}_{ij}\,\hat{\sigma}_i \otimes \hat{\sigma}_j,
\end{eqnarray}
where boldface variables $\bm{v}$ denote vectors, with $v = \vert \bm{v} \vert = \sqrt{\bm{v} \cdot \bm{v}} = \sqrt{\sum_{i=1}^{3} v_i^2}$, and
\begin{eqnarray}
\label{SecondH3}
\Tr[\mbox{\boldmath$\omega$}\cdot\mbox{\boldmath$\omega$}^T] &=& \omega_{ij} \omega_{ij},\\
\label{aaaaa}\mathcal{A}_i &=& 2 \beta_k\,\omega_{ik} = 2 (\mbox{\boldmath$\omega$} \cdot \mbox{\boldmath$\beta$})_i,\\
\mathcal{B}_j &=& 2 \alpha_k\,\omega_{kj} = 2 (\mbox{\boldmath$\alpha$} \cdot \mbox{\boldmath$\omega$})_j,\\
\label{ccccc}\mathcal{W}_{ij} &=& 2 \alpha_i\beta_j- \omega_{rs}\omega_{kl}\,\epsilon_{rki}\epsilon_{slj}.
\label{novod}
\end{eqnarray}
In particular, the general form of $\mathcal{W}_{ij}$, for unconstrained $\omega_{ij}$ parameters, can be manipulated in order to give
\begin{equation}
\label{SecondH3BB}
\mathcal{W}_{ij}=2 \left[\alpha_i\beta_j - \omega_{jk}\omega_{ki} + \omega_{ji}\Tr[\mbox{\boldmath$\omega$}]\right]-\delta_{ij}\left(\Tr[\mbox{\boldmath$\omega$}]^2-\Tr[\mbox{\boldmath$\omega$}^2]\right).
\end{equation}
However, it does not yield algebraic solutions for the Hamiltonian eigensystems without additional constraints over {\boldmath$\omega$}.
As an example, for $\omega_{ij}$ in the diagonal form, $\omega_{ij} = \omega_{(i)}\delta_{ij}$, one has $$2(\omega_{jk}\omega_{ki} - \omega_{ji}\Tr[\mbox{\boldmath$\omega$}])+\delta_{ij}\left(\Tr[\mbox{\boldmath$\omega$}]^2-\Tr[\mbox{\boldmath$\omega$}^2]\right)=0$$ at Eq.~\eqref{SecondH3BB}.
Similarly, identifying $\omega_{ij}$ by a dyadic form, $\omega_{ij}= a_i\,b_j$, yields $\omega_{rs}\omega_{kl}$ as a symmetric tensor under $r$ and $k$ permutation, i.e. $\omega_{rs}\omega_{kl}= a_r b_s\,a_k b_l = a_k b_s\,a_r b_l = \omega_{ks}\omega_{rl}$. Given that $\epsilon_{rki}$ is totally antisymmetric under index permutations, one thus has $\omega_{rs}\omega_{kl}\,\epsilon_{rki}\epsilon_{slj} = \omega_{ks}\omega_{rl}\,\epsilon_{rki} \epsilon_{slj}= 0$.
Both simplifications lead to algebraically solvable eigensystems.

\paragraph*{Case 01 -- Algorithm for separable states --} For the cases in which $\omega_{rs}\omega_{kl}\,\epsilon_{rki}\epsilon_{slj}$ are converted into a null contribution, $\mathcal{W}_{ij}$ is reduced to
\begin{eqnarray}
\label{SecondH3B}
\mathcal{W}_{ij} &=& 2 \alpha_i\beta_j.
\end{eqnarray}
In this case, by repeating the above algorithm over the operator $\tilde{H}^2$, one straightforwardly obtains another traceless operator,
\begin{equation}
\label{ThirdH}
\mathcal{O} = \tilde{H}^2 - \mathcal{V} \,\hat{I}_{_{{ 4}}},
\end{equation}
with
\begin{equation}
\label{ThirdH2}
\mathcal{V} = \frac{1}{4}\Tr[\tilde{H}^2 ] = \mbox{\boldmath$\alpha$}^2 + \mbox{\boldmath$\beta$}^2 + \Tr[\mbox{\boldmath$\omega$}\cdot\mbox{\boldmath$\omega$}^T],
\end{equation}
from which one analogously identifies
\begin{eqnarray}
\label{ThirdH3}
\mathcal{O}^2 &=& \left\{\mbox{\boldmath$\mathcal{A}$}^2 + \mbox{\boldmath$\mathcal{B}$}^2 + \Tr[\mbox{\boldmath$\mathcal{W}$}\cdot\mbox{\boldmath$\mathcal{W}$}^T] \right\} \hat{I}_{_{{ 2}}} \otimes\hat{I}_{_{{ 2}}}\nonumber\\
&&\qquad+2\,\mathcal{B}_s\mathcal{W}_{is}\, \hat{\sigma}_i \otimes\hat{I}_{_{{ 2}}} 
+2\,\mathcal{A}_s\mathcal{W}_{sj}\,\hat{I}_{_{{ 2}}} \otimes\hat{\sigma}_j
+2\,\mathcal{A}_i\mathcal{B}_j\,\hat{\sigma}_i \otimes \hat{\sigma}_j.\qquad
\end{eqnarray}

Thus, by noticing that 
\begin{eqnarray}
\label{ThirdH4}
\mbox{\boldmath$\mathcal{A}$}^2&=& 4\,\mbox{\boldmath$\alpha$}\cdot\mbox{\boldmath$\omega$}\cdot\mbox{\boldmath$\omega$}^{T}\hspace{-.15cm}\cdot\mbox{\boldmath$\alpha$}\\
\mbox{\boldmath$\mathcal{B}$}^2&=& 4\,\mbox{\boldmath$\beta$}\cdot\mbox{\boldmath$\omega$}^{T}\hspace{-.15cm}\cdot\mbox{\boldmath$\omega$}\cdot\mbox{\boldmath$\beta$},\\
\Tr[\mbox{\boldmath$\mathcal{W}$}\cdot\mbox{\boldmath$\mathcal{W}$}^T] &=& 4\,\mbox{\boldmath$\alpha$}^2\mbox{\boldmath$\beta$}^2,\\
\mathcal{A}_s \mathcal{W}_{sj} &=& 
4\beta_k \omega_{sk} \alpha_s \beta_j = 4 (\alpha_s \omega_{sk} \beta_k)\beta_j = 4\, (\mbox{\boldmath$\alpha$}\cdot\mbox{\boldmath$\omega$}\cdot\mbox{\boldmath$\beta$})\,\beta_j,\\
\mathcal{B}_s \mathcal{W}_{is} &=& 
4\alpha_k \omega_{ks} \alpha_i \beta_s = 4 (\alpha_k \omega_{ks} \beta_s)\alpha_i = 4\, (\mbox{\boldmath$\alpha$}\cdot\mbox{\boldmath$\omega$}\cdot\mbox{\boldmath$\beta$})\,\alpha_i,\\
\mathcal{A}_i\mathcal{B}_j &=&
4\beta_k \omega_{\underline{i}k} \alpha_s \omega_{\underline{s}j} = 4 (\alpha_s \omega_{sk} \beta_k)\omega_{ij} = 4 \,(\mbox{\boldmath$\alpha$}\cdot\mbox{\boldmath$\omega$}\cdot\mbox{\boldmath$\beta$})\,\omega_{ij},
\end{eqnarray}
once they are substituted into Eq.~\eqref{ThirdH3}, the operator $\mathcal{O}^2$ is reduced to a linear dependence on $\tilde{H}$, 
\begin{eqnarray}
\label{ThirdH3B}
\mathcal{O}^2 &=& \tilde{H}^4 - 2\mathcal{V} \tilde{H}^2 + \mathcal{V}^2 \hat{I}_{_{{ 4}}}\nonumber\\
&=& 4\left\{\mbox{\boldmath$\alpha$}\cdot\mbox{\boldmath$\omega$}\cdot\mbox{\boldmath$\omega$}^{T}\hspace{-.15cm}\cdot\mbox{\boldmath$\alpha$}+\mbox{\boldmath$\beta$}\cdot\mbox{\boldmath$\omega$}^{T}\hspace{-.15cm}\cdot\mbox{\boldmath$\omega$}\cdot\mbox{\boldmath$\beta$}+\mbox{\boldmath$\alpha$}^2\mbox{\boldmath$\beta$}^2\right\}\hat{I}_{_{{ 4}}} +8\,(\mbox{\boldmath$\alpha$}\cdot\mbox{\boldmath$\omega$}\cdot\mbox{\boldmath$\beta$}) \tilde{H},
\end{eqnarray} 
which constrains the quartic powers of $\tilde{H}$ to a linear combination of its lower powers: $\tilde{H}^2$, $\tilde{H}$ and $\tilde{H}^0\equiv \hat{I}_{_{{ 4}}}$.
Hence, the operator $\mathcal{O}^2$, once applied to $\rho_{mn}$ states, through the identification of the eigensystem,
\begin{equation}
\label{FirstH2B}
\tilde{H} \,\rho_{mn} = \mathcal{E}_{mn} \hat{I}_{_{{ 4}}}^{(1\otimes 2)} \, \rho_{mn},
\end{equation}
yields the quartic secular equation for the associated eigenvalues, 
\begin{equation}
\label{ForthH}
\mathcal{E}^4 - 2 \mathcal{E}^2 \,\mathcal{V} - 8\mathcal{E}\, (\mbox{\boldmath$\alpha$}\cdot\mbox{\boldmath$\omega$}\cdot\mbox{\boldmath$\beta$}) - \Theta + \mathcal{V}^2=0,
\end{equation}
with
\begin{equation}
\label{ForthHSS}
\Theta = \frac{1}{4} \Tr[\mathcal{O}^2] = 4\left\{\mbox{\boldmath$\alpha$}\cdot\mbox{\boldmath$\omega$}\cdot\mbox{\boldmath$\omega$}^{T}\hspace{-.15cm}\cdot\mbox{\boldmath$\alpha$}+\mbox{\boldmath$\beta$}\cdot\mbox{\boldmath$\omega$}^{T}\hspace{-.15cm}\cdot\mbox{\boldmath$\omega$}\cdot\mbox{\boldmath$\beta$}+\mbox{\boldmath$\alpha$}^2\mbox{\boldmath$\beta$}^2\right\},\end{equation}
and $\mathcal{E}$ identified by $\mathcal{E}_{mn} = \varepsilon_{mn} - \upsilon$.
Eq.~\eqref{ForthH} admits exact algebraic solutions by means of the Cardano-Ferrari's method \cite{Cardano}.
However, obtaining the eigenvectors, $\rho_{mn}$, through some general algorithm, corresponds to a more challenging issue.
As it shall be addressed in the following, for separable states, $\rho_{mn} \equiv \rho_m^{(1)} \otimes \rho_n^{(2)}$, the above quartic equation can be decoupled into simplest quadratic equations which yield an algebraically solvable eigensystem.

\paragraph*{Case 02 -- Algorithm for entangled states --} For non-separable eigenstates, $\rho_{mn} \equiv \rho_{mn}^{(1\otimes2)}$, one should turn back to the general form of $\mathcal{W}_{ij}$, Eq.~\eqref{SecondH3BB}, and introduce an additional constraint either with $\mbox{\boldmath$\alpha$}\cdot\mbox{\boldmath$\omega$}=0$ or with $\mbox{\boldmath$\omega$}\cdot\mbox{\boldmath$\beta$}=0$, to solve the eigensystem algebraically. One would have $\mathcal{A}_i\mathcal{B}_j =0$ and either
$$\mbox{\boldmath$\mathcal{A}$}^2= 4\,\mbox{\boldmath$\beta$}\cdot\mbox{\boldmath$\omega$}\cdot\mbox{\boldmath$\omega$}^{T}\hspace{-.15cm}\cdot\mbox{\boldmath$\beta$} \quad \mbox{and}\quad \mathcal{B}_s \mathcal{W}_{is} =\mbox{\boldmath$\mathcal{B}$}^2=0;$$
or
$$\mbox{\boldmath$\mathcal{B}$}^2 = 4\,\mbox{\boldmath$\alpha$}\cdot\mbox{\boldmath$\omega$}^{T}\hspace{-.15cm}\cdot\mbox{\boldmath$\omega$}\cdot\mbox{\boldmath$\alpha$}\quad\mbox{and}\quad \mathcal{A}_s \mathcal{W}_{sj} =\mbox{\boldmath$\mathcal{A}$}^2=0,$$
respectively.

Furthermore,
\footnotesize\begin{eqnarray}\label{above}
\Phi=\Tr[\mbox{\boldmath$\mathcal{W}$}\cdot\mbox{\boldmath$\mathcal{W}$}^T] &=& 4\left\{\mbox{\boldmath$\alpha$}^2\mbox{\boldmath$\beta$}^2 - \mbox{\boldmath$\alpha$}\cdot\mbox{\boldmath$\beta$}\left(\Tr[\mbox{\boldmath$\omega$}]^2-\Tr[\mbox{\boldmath$\omega$}^2]\right)+ \Tr[\mbox{\boldmath$\omega$}^4]\right\}\nonumber\\&& - 4\Tr[\mbox{\boldmath$\omega$}^3] \Tr[\mbox{\boldmath$\omega$}] + 4\Tr[\mbox{\boldmath$\omega$}^2]\Tr[\mbox{\boldmath$\omega$}]^2- \left(\Tr[\mbox{\boldmath$\omega$}]^2-\Tr[\mbox{\boldmath$\omega$}^2]\right)^2\nonumber\\
 &=& 4\left\{\mbox{\boldmath$\alpha$}^2\mbox{\boldmath$\beta$}^2 - \mbox{\boldmath$\alpha$}\cdot\mbox{\boldmath$\beta$}\left(\Tr[\mbox{\boldmath$\omega$}]^2-\Tr[\mbox{\boldmath$\omega$}^2]\right)+ \Tr[\mbox{\boldmath$\omega$}^4]\right\}\nonumber\\&& - 2\Tr[\mbox{\boldmath$\omega$}]\left\{2\Tr[\mbox{\boldmath$\omega$}^3] - 3\Tr[\mbox{\boldmath$\omega$}^2]\Tr[\mbox{\boldmath$\omega$}]+\Tr[\mbox{\boldmath$\omega$}]^3\right\}+ \Tr[\mbox{\boldmath$\omega$}]^4 - \Tr[\mbox{\boldmath$\omega$}^2]^2\nonumber\\
 &=& 4\left\{\mbox{\boldmath$\alpha$}^2\mbox{\boldmath$\beta$}^2 - \mbox{\boldmath$\alpha$}\cdot\mbox{\boldmath$\beta$}\left(\Tr[\mbox{\boldmath$\omega$}]^2-\Tr[\mbox{\boldmath$\omega$}^2]\right)+ \Tr[\mbox{\boldmath$\omega$}^4]\right\}
\nonumber\\&&
- 12\Tr[\mbox{\boldmath$\omega$}]\,\det[\mbox{\boldmath$\omega$}]+ \Tr[\mbox{\boldmath$\omega$}]^4 - \Tr[\mbox{\boldmath$\omega$}^2]^2\\
 &=& 4\left\{\mbox{\boldmath$\alpha$}^2\mbox{\boldmath$\beta$}^2 - \left[\mbox{\boldmath$\alpha$}\cdot\mbox{\boldmath$\beta$}-\frac{\Tr[\mbox{\boldmath$\omega$}]^2 + \Tr[\mbox{\boldmath$\omega$}^2]}{4}\right]\left(\Tr[\mbox{\boldmath$\omega$}]^2-\Tr[\mbox{\boldmath$\omega$}^2]\right)+ \Tr[\mbox{\boldmath$\omega$}^4]\right\} - 12\Tr[\mbox{\boldmath$\omega$}]\,\det[\mbox{\boldmath$\omega$}]\nonumber
\end{eqnarray}\normalsize
where, in the third step, the Cayley-Hamilton theorem has been used.
The above expression, Eq.~\eqref{above}, should replace the $4\mbox{\boldmath$\alpha$}^2\mbox{\boldmath$\beta$}^2$ contribution in $\Theta$, Eq.~\eqref{ForthHSS}, in order to set the operator $\mathcal{O}^2$ reduced to
\begin{eqnarray}
\label{ThirdH3Bmm}
\mathcal{O}^2 &=& \tilde{H}^4 - 2\mathcal{V} \tilde{H}^2 + \left(\mathcal{V}^2-\Theta_{_{\Phi}}\right) \hat{I}_{_{{ 4}}}
\end{eqnarray}
with 
\begin{equation}
\label{ForthHSSB}
\Theta_{_{\Phi}} = 4\left\{\delta_{{\tiny{\mbox{\boldmath$\omega$}\cdot\mbox{\boldmath$\beta$}}}}^{0}\left(\mbox{\boldmath$\alpha$}\cdot\mbox{\boldmath$\omega$}\cdot\mbox{\boldmath$\omega$}^{T}\hspace{-.15cm}\cdot\mbox{\boldmath$\alpha$}\right)+\delta_{{\tiny{\mbox{\boldmath$\alpha$}\cdot\mbox{\boldmath$\omega$}}}}^{0}\left(\mbox{\boldmath$\beta$}\cdot\mbox{\boldmath$\omega$}^{T}\hspace{-.15cm}\cdot\mbox{\boldmath$\omega$}\cdot\mbox{\boldmath$\beta$}\right)\right\}+\Phi,\end{equation}

Further simplifications can be introduced if one pays attention to the choice of the frame system. 
Setting $\mbox{\boldmath$\alpha$}\cdot\mbox{\boldmath$\omega$}=0$ (or equivalently $\mbox{\boldmath$\beta$}\cdot\mbox{\boldmath$\omega$}=0$) allows for recasting $\mbox{\boldmath$\omega$}$ into a two-dimensional symmetric block diagonal form, $\mbox{\boldmath$\omega$}_{_B}$, such that
\begin{eqnarray}
\label{above4}
\det[\mbox{\boldmath$\omega$}]&=&0,\nonumber\\
\Tr[\mbox{\boldmath$\omega$}]^2-\Tr[\mbox{\boldmath$\omega$}^2]&=&2\det[\mbox{\boldmath$\omega$}_{_B}],\nonumber
\end{eqnarray}
and
\begin{eqnarray}
\label{ajuda4}
2(\omega_{jk}\omega_{ki} - \omega_{ji}\Tr[\mbox{\boldmath$\omega$}])+\delta_{ij}\left(\Tr[\mbox{\boldmath$\omega$}]^2-\Tr[\mbox{\boldmath$\omega$}^2]\right)&=& 2\det[\mbox{\boldmath$\omega$}_{_B}]\,
\left(\delta_{{\tiny{\mbox{\boldmath$\omega$}\cdot\mbox{\boldmath$\beta$}}}}^{0}\frac{\beta_i\beta_j}{\mbox{\boldmath$\beta$}^2}+
\delta_{{\tiny{\mbox{\boldmath$\alpha$}\cdot\mbox{\boldmath$\omega$}}}}^{0}\frac{\alpha_i\alpha_j}{\mbox{\boldmath$\alpha$}^2}
\right),\nonumber
\end{eqnarray}
which leads to
\begin{equation}
\label{SecondH3BB22}
\mathcal{W}_{ij}=2 \left\{\alpha_i\beta_j - \det[\mbox{\boldmath$\omega$}_{_B}]\,\left(\delta_{{\tiny{\mbox{\boldmath$\omega$}\cdot\mbox{\boldmath$\beta$}}}}^{0}\frac{\beta_i\beta_j}{\mbox{\boldmath$\beta$}^2}+
\delta_{{\tiny{\mbox{\boldmath$\alpha$}\cdot\mbox{\boldmath$\omega$}}}}^{0}\frac{\alpha_i\alpha_j}{\mbox{\boldmath$\alpha$}^2}
\right)\right\},
\end{equation}
$\mathcal{B}_s \mathcal{W}_{is}= \mathcal{A}_s \mathcal{W}_{sj} = 0$, and
\begin{eqnarray}
\label{above2}
\Phi
 &=& 4\left(
 \mbox{\boldmath$\alpha$}^2\mbox{\boldmath$\beta$}^2 + \det[\mbox{\boldmath$\omega$}_{_B}]^2 - 2\mbox{\boldmath$\alpha$}\cdot\mbox{\boldmath$\beta$}\det[\mbox{\boldmath$\omega$}_{_B}]\right)\nonumber\\
 &=& 4\left[\left(\mbox{\boldmath$\alpha$}\cdot\mbox{\boldmath$\beta$}-
 \det[\mbox{\boldmath$\omega$}_{_B}]\right)^2+ (\mbox{\boldmath$\alpha$}\times\mbox{\boldmath$\beta$})^2\right],
 \end{eqnarray}
with \small\begin{eqnarray}
\label{ForthHSSBFim}
\Theta_{_{\Phi}} &=& 4\left\{\delta_{{\tiny{\mbox{\boldmath$\beta$}\cdot\mbox{\boldmath$\omega$}}}}^{0}\left(\mbox{\boldmath$\alpha$}\cdot\mbox{\boldmath$\omega$}\cdot\mbox{\boldmath$\omega$}^{T}\hspace{-.15cm}\cdot\mbox{\boldmath$\alpha$}\right)+\delta_{{\tiny{\mbox{\boldmath$\alpha$}\cdot\mbox{\boldmath$\omega$}}}}^{0}\left(\mbox{\boldmath$\beta$}\cdot\mbox{\boldmath$\omega$}^{T}\hspace{-.15cm}\cdot\mbox{\boldmath$\omega$}\cdot\mbox{\boldmath$\beta$}\right)+\left(\mbox{\boldmath$\alpha$}\cdot\mbox{\boldmath$\beta$}-
 \det[\mbox{\boldmath$\omega$}_{_B}]\right)^2+ (\mbox{\boldmath$\alpha$}\times\mbox{\boldmath$\beta$})^2\right\}.\nonumber\\
 &=& 4\left\{\delta_{{\tiny{\mbox{\boldmath$\beta$}\cdot\mbox{\boldmath$\omega$}}}}^{0}
\left[\mbox{\boldmath$\alpha$}\cdot\mbox{\boldmath$\omega$}\cdot\mbox{\boldmath$\alpha$}\,\Tr[\mbox{\boldmath$\omega$}] + \det[\mbox{\boldmath$\omega$}_{_B}]\left((\mbox{\boldmath$\alpha$}\cdot\mbox{\boldmath$\beta$})^2\mbox{\boldmath$\beta$}^{-2}\right)-\mbox{\boldmath$\alpha$}^2 \right]\right.\nonumber\\
&&\qquad\qquad
+\delta_{{\tiny{\mbox{\boldmath$\alpha$}\cdot\mbox{\boldmath$\omega$}}}}^{0}
\left[\mbox{\boldmath$\beta$}\cdot\mbox{\boldmath$\omega$}\cdot\mbox{\boldmath$\beta$}\,\Tr[\mbox{\boldmath$\omega$}]+ \det[\mbox{\boldmath$\omega$}_{_B}]\left( (\mbox{\boldmath$\alpha$}\cdot\mbox{\boldmath$\beta$})^2\mbox{\boldmath$\alpha$}^{-2}\right)-\mbox{\boldmath$\beta$}^2 \right]\nonumber\\
&&\qquad\qquad\qquad\qquad\qquad\qquad\qquad\left.
+\left(\mbox{\boldmath$\alpha$}\cdot\mbox{\boldmath$\beta$}-
 \det[\mbox{\boldmath$\omega$}_{_B}]\right)^2+ (\mbox{\boldmath$\alpha$}\times\mbox{\boldmath$\beta$})^2\right\},\end{eqnarray}\normalsize
since, from Eq.~\eqref{ajuda4},
\begin{eqnarray}
\label{ajuda3}
\omega_{jk}\omega_{ki} = \omega_{ji}\Tr[\mbox{\boldmath$\omega$}]+
 \det[\mbox{\boldmath$\omega$}_{_B}]\left(\delta_{{\tiny{\mbox{\boldmath$\omega$}\cdot\mbox{\boldmath$\beta$}}}}^{0}\frac{\beta_i\beta_j}{\mbox{\boldmath$\beta$}^2}+
\delta_{{\tiny{\mbox{\boldmath$\alpha$}\cdot\mbox{\boldmath$\omega$}}}}^{0}\frac{\alpha_i\alpha_j}{\mbox{\boldmath$\alpha$}^2}
-\delta_{ij}\right).\nonumber
\end{eqnarray}

\subsection{Separable states}

The dynamics driven by $\hat{H}$ would be considerably simplified if the Hilbert space $\mathbb{H} = \mathbb{H}^{(1)} \otimes \mathbb{H}^{(2)}$ could be reduced to the subspace $\mathbb{H}^{(1)} \oplus \mathbb{H}^{(2)}$. In this case, the Hamiltonian from Eq.~\eqref{FirstHcc} would be reduced to
\begin{equation}
\label{FirstHcc2}
\hat{H}_S \equiv \hat{H}^{(1)}\otimes \hat{H}^{(2)},
\end{equation}
with $H^{(1)}$ and $H^{(2)}$ written as $H^{(1)} = a_{\mu} \hat{\sigma}_{\mu}$ and $H^{(2)} = b_{\mu} \hat{\sigma}_{\mu}$, with $\mu = 0,\,1,\,2,\,3$, and $\hat{\sigma}_0 = \hat{I}_{_{{ 2}}}$. 

From the correspondence with Eq.~\eqref{FirstH}, identifying $\omega_{ij}$ by the dyadic form, $\omega_{ij}= a_i\,b_j$, suppresses the needs of an additional constraint over $\mathcal{W}_{ij}$ from Eq.~\eqref{novod}, which is straightforwardly reduced to $\mathcal{W}_{ij} = 2 \alpha_i \beta_j$ from Eq.~\eqref{SecondH3B}.

Besides $\omega_{ij} = a_i\,b_j$, the following constraints are identified: $\upsilon = a_0\,b_0$, $\alpha_i = -b_0 \,a_i$, and $\beta_i = -a_0 \,b_i$. Once they are substituted into the quartic secular equation, Eq.~\eqref{ForthH}, it can be recasted as the product of two simpler quadratic equations\footnote{In correspondence with
\begin{eqnarray}
\label{ForthH3}
\left(\hat{H}^{(1)}\right)^2 &=& a_{\mu} a_{\nu} \,\hat{\sigma}_{\nu}\hat{\sigma}_{\mu} = 2\,a_0\hat{H}^{(1)} -(a_0^2 - a^2)\hat{I}_{_{{ 2}}},\\
\left(\hat{H}^{(2)}\right)^2 &=& b_{\mu} b_{\nu} \,\hat{\sigma}_{\nu}\hat{\sigma}_{\mu} = 2\,b_0\hat{H}^{(2)} - (b_0^2 - b^2)\hat{I}_{_{{ 2}}},
\end{eqnarray}
respectively.},
\begin{eqnarray}
\label{ForthH2}
\varepsilon^2_m - 2 \varepsilon_m\,a_0+ (a_0^2 - a^2) &=&0,\\
\varepsilon^2_n - 2 \varepsilon_n\,b_0+ (b_0^2 - b^2) &=&0,
\end{eqnarray}
with $m,\,n = 1,\,2$, and from which one obtains
\begin{eqnarray}
\label{ForthH2}
\varepsilon_m &=& a_0 + (-1)^m a,\\
\varepsilon_n &=& b_0 + (-1)^n b,
\end{eqnarray}
which lead to the $\hat{H}_S$ eigenvalues
\begin{eqnarray}
\label{ForthH2B}
\varepsilon_{mn} = \varepsilon_m\varepsilon_n = (a_0 + (-1)^m a)(b_0 + (-1)^n b).
\end{eqnarray}
Of course, due to the nature of the Hamiltonian $\hat{H}_S$, the eigenstates associated with $\varepsilon_{mn}$ are separable, and can be cast into the form of $\rho_{mn} \equiv \rho_m^{(1)} \otimes \rho_n^{(2)}$, with
\begin{eqnarray}
\label{ForthH2}
\label{111}\rho_m^{(1)} &=& 
\frac{1}{2}\left[\hat{I}_{_{{ 2}}} + \frac{\hat{H}^{(1)}-a_0\hat{I}_{_{{ 2}}}}{\varepsilon_m-a_0}\right]
=\frac{1}{2}\left[\hat{I}_{_{{ 2}}} +(-1)^m \frac{a_i \,\hat{\sigma}_i}{a}\right],\\
\label{222}\rho_n^{(2)} &=& 
\frac{1}{2}\left[\hat{I}_{_{{ 2}}} + \frac{\hat{H}^{(2)}-b_0\hat{I}_{_{{ 2}}}}{\varepsilon_n-b_0}\right]
=\frac{1}{2}\left[\hat{I}_{_{{ 2}}} +(-1)^n \frac{b_i \,\hat{\sigma}_i}{b}\right].
\end{eqnarray}
From the above expressions, Eqs.~\eqref{111} and \eqref{222}, it is straightforward to verify that $\{\rho_{mn}\}$, with $m,\,n = 1,\,2$, form an orthonormal basis of pure states (i.e. $\Tr[\rho_{mn}]=1$ and $\Tr[\rho_{mn}\rho_{pq}] = \delta_{mp}\delta_{nq}$). Despite its simple form, it is interesting to notice that $\rho_{mn}$ cannot be expressed exclusively in terms of powers of $\tilde{H}_S$, the corresponding traceless component of the Hamiltonian $\hat{H}_S$ from Eq.~\eqref{FirstHcc2}.
As it shall be discussed in the following, a stratagem for obtaining the $SU(2)\otimes SU(2)$ Hamiltonian eigenstate basis in terms of powers of traceless operators, $\tilde{H}$, can be relevant for solving the generic eigensystem problem, Eq.~\eqref{FirstH2}.

\subsubsection{TD ensembles}

All the necessary elements for constructing the partition function, $\mathcal{Z}(T)$, for the above discussed quantum system were provided. The thermal density matrix is identified by
\begin{equation}\label{therma}
\varrho(T) = \frac{1}{\mathcal{Z}(T)}\sum_{m,n=1,2}\rho_{mn} \, \exp[-\varepsilon_{mn}/k_{_B} T],
\end{equation}
where $k_{_B}$ is the Boltzmann constant, and $T$ is the ensemble temperature.
From Eqs.~\eqref{ForthH2B} and \eqref{ForthH2}, one thus has
\begin{eqnarray}
\varrho(T) &=& \frac{1}{4\mathcal{Z}(T)}\sum_{m,n=1,2}\exp[-(a_0 + (-1)^m a)(b_0 + (-1)^n b)/k_{_B} T]\nonumber\\
&&\qquad\qquad\qquad\qquad \times\left\{\left[\hat{I}_{_{{ 2}}} +(-1)^m \frac{a_i \,\hat{\sigma}_i}{a}\right]
\otimes
\left[\hat{I}_{_{{ 2}}} +(-1)^n \frac{b_i \,\hat{\sigma}_i}{b}\right]\right\},
\end{eqnarray}
which, from straightforward manipulations, results in\footnote{It has been used that
\begin{eqnarray}
\exp\left(-\tilde{H}^{(1)}b_0/k_{_B} T\right)\otimes \exp\left(-\tilde{H}^{(2)}a_0/k_{_B} T\right) &=& 
\exp\left(-\tilde{H}^{(1)}b_0/k_{_B} T\right)\otimes \hat{I}_{_{{ 2}}}\times\hat{I}_{_{{ 2}}}\otimes \exp\left(-\tilde{H}^{(2)}a_0/k_{_B} T\right)\nonumber\\
&=& \exp\left(-\tilde{H}^{(1)}b_0/k_{_B} T\otimes \hat{I}_{_{{ 2}}}\right)\times \exp\left(-\hat{I}_{_{{ 2}}}\otimes\tilde{H}^{(2)}a_0/k_{_B} T\right).\nonumber\end{eqnarray}}

\begin{eqnarray}
\varrho(T) &=& \frac{e^{-\frac{a_0 b_0}{k_{_B} T}}}{\mathcal{Z}(T)}
\left\{
\cosh(a b /k_{_B} T)\left[\cosh(a b_0 /k_{_B} T)\hat{I}_{_{{ 2}}} -\sinh(a b_0 /k_{_B} T) \frac{a_i \,\hat{\sigma}_i}{a}\right]
\right.\nonumber\\
&&\qquad\qquad\qquad\qquad\qquad\qquad\otimes
\left[\cosh(a_0 b /k_{_B} T)\hat{I}_{_{{ 2}}} -\sinh(a_0 b /k_{_B} T) \frac{b_i \,\hat{\sigma}_i}{b}\right]
\nonumber\\
&&\qquad\qquad
-\sinh(a b /k_{_B} T)\left[\sinh(a b_0 /k_{_B} T)\hat{I}_{_{{ 2}}} -\cosh(a b_0 /k_{_B} T) \frac{a_i \,\hat{\sigma}_i}{a}\right]
\nonumber\\
&&\qquad\qquad\qquad\qquad\qquad\qquad\qquad\left.
\otimes\left[\sinh(a_0 b /k_{_B} T)\hat{I}_{_{{ 2}}} -\cosh(a_0 b /k_{_B} T) \frac{b_i \,\hat{\sigma}_i}{b}\right]\right\}\nonumber\\
&=&\frac{e^{-\frac{a_0 b_0}{k_{_B} T}}}{\mathcal{Z}(T)}
\left\{
\cosh(a b /k_{_B} T)
\left[\exp\left(-\tilde{H}^{(1)}b_0/k_{_B} T\right)\otimes \exp\left(-\tilde{H}^{(2)}a_0/k_{_B} T\right)\right]\right.\nonumber\\
&&\qquad\qquad\left.
-\sinh(a b /k_{_B} T)
\left[{\tilde{H}^{(1)}}{a}^{-1}\exp\left(-\tilde{H}^{(1)}b_0/k_{_B} T\right)\otimes {\tilde{H}^{(2)}}{b}^{-1}\exp\left(-\tilde{H}^{(2)}a_0/k_{_B} T\right)\right]\right\}
\nonumber\\
&=&\frac{e^{-\frac{a_0 b_0}{k_{_B} T}}}{\mathcal{Z}(T)}\left[\exp\left(-\tilde{H}^{(1)}b_0/k_{_B} T\right)\otimes \exp\left(-\tilde{H}^{(2)}a_0/k_{_B} T\right)\right]
\nonumber\\
&&\qquad\qquad\left\{
\cosh(a b /k_{_B} T)
-\sinh(a b /k_{_B} T)
\left[{\tilde{H}^{(1)}}{a}^{-1}\otimes {\tilde{H}^{(2)}}{b}^{-1}\right]\right\}
\nonumber\\
&=&\frac{e^{-\frac{a_0 b_0}{k_{_B} T}}}{\mathcal{Z}(T)}\left[\exp\left(-\tilde{H}^{(1)}b_0/k_{_B} T\right)\otimes \exp\left(-\tilde{H}^{(2)}a_0/k_{_B} T\right)\right] \exp\left[-\left(\tilde{H}^{(1)}\otimes\tilde{H}^{(2)}\right)/k_{_B} T\right]
\nonumber\\
&=&
\exp\left[-\left(a_0 b_0 \hat{I}_{_{{ 2}}}\otimes \hat{I}_{_{{ 2}}}+\tilde{H}^{(1)}b_0\otimes \hat{I}_{_{{ 2}}} + \hat{I}_{_{{ 2}}}\otimes\tilde{H}^{(2)}a_0+\tilde{H}^{(1)}\otimes\tilde{H}^{(2)}\right)/k_{_B} T\right]/{\mathcal{Z}(T)}
\nonumber\\
&\equiv& \exp\left(-\hat{H}^{(1)}\otimes \hat{H}^{(2)}/k_{_B} T\right)/\mathcal{Z}(T),\label{therma}\qquad
\end{eqnarray}
from which the trace operation, $\Tr[\varrho(T)]=1$, leads to the partition function,
\begin{eqnarray}
\label{PF}
\mathcal{Z}(T) &=& \Tr\left[\exp\left(-\hat{H}^{(1)}\otimes \hat{H}^{(2)}/k_{_B} T\right)\right]\nonumber\\
&=& \sum_{m,n=1,2}\exp[-(a_0 + (-1)^m a)(b_0 + (-1)^n b)/k_{_B} T]
\nonumber\\
&=& 4\exp\left(-{a_0 b_0}/{k_{_B} T}\right)\left[\prod_{p=1}^3 \cosh(m_p/k_{_B} T)-\prod_{p=1}^3 \sinh(m_p/k_{_B} T)\right],
\end{eqnarray}
where $m_1=ab$, $m_2=a_0b$, and $m_3=ab_0$, and it has been assumed that the eigenstates are non-degenerate. 

Through the partition function, Eq.~\eqref{PF}, the associated quantum purity can be computed and expressed by 
\cite{Bernardini2020AA}
\begin{eqnarray}
\label{PF}
\mathcal{P}(T) &=&\frac{\mathcal{Z}(T/2)}{(\mathcal{Z}(T))^2}
\nonumber\\
&=&\frac{1}{4} {\prod_{p=1}^3 \left[1+\tanh^2(m_p/k_{_B} T)\right]\,\left[1-\prod_{p=1}^3 \tanh(2m_p/k_{_B} T)\right]}{\left[1-\prod_{p=1}^3 \tanh(m_p/k_{_B} T)\right]^{-2}}
\quad
\end{eqnarray}
which is consistent with the classical and pure state limits, $\mathcal{P}(\infty) = 1/4$ and $\mathcal{P}(0) = 1$, respectively.
Finally, bipartite states $\rho$, as those described either by $\rho_{mn}$ or by $\varrho(T)$, are entangled if they cannot be expressed as \cite{Entanglement01}
\begin{equation}\label{bipar}
\rho = \displaystyle \sum_i \eta_i \,\rho_i^{(1)} \otimes \rho_i^{(2)},
\end{equation}
where $\rho^{(1)}_i \in H_1$, $\rho^{(2)}_i \in H_2$, and $ \sum_i \eta_i = 1$. Of course, $\rho_{mn} \equiv \rho_m^{(1)} \otimes \rho_n^{(2)}$ are separable. Likewise, from Eq.~\eqref{therma}, $\varrho(T)$ can be read from the above expansion,
\begin{equation}
\varrho(T) = \frac{1}{\mathcal{Z}(T)}\sum_{m,n=1,2}\rho_m^{(1)} \otimes \rho_n^{(2)}\, \exp[-\varepsilon_{mn}/k_{_B} T],
\end{equation}
as a separable state.

\subsection{Entangled pure states}

Turning back to the traceless Hamiltonian, $\tilde{H}$, in the form resumed by Eqs.~\eqref{FirstH3} and \eqref{SecondH}, with $\mathcal{W}_{ij}$ from Eq.~\eqref{SecondH3BB22}, one notices that
\begin{equation}
\Tr[\tilde{H}^3] \propto \mbox{\boldmath$\alpha$}\cdot\mbox{\boldmath$\omega$}\cdot\mbox{\boldmath$\beta$}.
\end{equation}
Analogously, by multiplying Eqs.~\eqref{ThirdH} and \eqref{ThirdH3B} by $\tilde{H}$, with $\Tr[\tilde{H}]=0$, one also has, respectively,
\begin{eqnarray}
\label{Fifth}
\Tr[\tilde{H}\mathcal{O}] &=&\Tr[\tilde{H}^3] \propto \mbox{\boldmath$\alpha$}\cdot\mbox{\boldmath$\omega$}\cdot\mbox{\boldmath$\beta$},\\
\Tr[\tilde{H}\mathcal{O}^2]\Rightarrow \Tr[\tilde{H}^5]&=& 2 \mathcal{V} \Tr[\tilde{H}^3] \propto \mbox{\boldmath$\alpha$}\cdot\mbox{\boldmath$\omega$}\cdot\mbox{\boldmath$\beta$}.
\end{eqnarray}
This means that including an additional constraint, $\mbox{\boldmath$\alpha$}\cdot\mbox{\boldmath$\omega$}\cdot\mbox{\boldmath$\beta$}=0$, besides introducing an even symmetry to the eigenvalue Eq.~\eqref{ForthH}, now recast as
\begin{equation}
\label{ForthHBBB}
\left(\mathcal{E}^2 - \mathcal{V} \right)^2 = \Theta_{_{\Phi}},
\end{equation}
constrains the odd powers of $\tilde{H}$ to
\begin{equation}
\Tr[\tilde{H}^{2N+1}] = \Tr[\tilde{H}\mathcal{O}] = 0, \qquad \mbox{with $N$ integer.}
\end{equation}
In this case, the Hamiltonian eigenvalues are then identified by
\begin{eqnarray}
\label{ForthH2BNN}
\varepsilon_{mn}= \mathcal{E}_{mn}+\upsilon = \upsilon + (-1)^m \left[\mathcal{V} + (-1)^n \Theta_{_{\Phi}}^{\frac{1}{2}}\right]^{\frac{1}{2}},
\end{eqnarray}
with a straightforward correspondence with pure eigenstates given by the following {\em ansatz}\footnote{It is relevant to notice that the hypothesis of vanishing denominators at Eq.~\eqref{ForthH2NN}, namely for either $ \Theta_{_{\Phi}}=0$ or $\mathcal{V} + (-1)^n \Theta_{_{\Phi}}^{\frac{1}{2}}=0$ does not happens without equivalent conditions for the corresponding numerators. Firstly, $\Theta_{_{\Phi}}=0$ is only possible for trivial values of the coefficients defined by Eqs.~\eqref{aaaaa}-\eqref{ccccc}, which implies that $\tilde{H} = \mathcal{V}\hat{I}_{_{4}}$ and, therefore, $\mathcal{O} \sim (-1)^n \Theta_{_{\Phi}}^{\frac{1}{2}} \hat{I}_{_{4}}$, with the resulting related coefficient at Eq.~\eqref{ForthH2NN} canceling the denominator. Secondly, and more drastically, 
$\mathcal{V} + (-1)^n \Theta_{_{\Phi}}^{\frac{1}{2}}=0$ is only possible for simultaneous vanishing values of $\alpha_i$, $\beta_i$, and $\omega_{ij}$, which implies that $\hat{H} = \upsilon\hat{I}_{_{4}}$ and, therefore, $\tilde{H} \sim (-1)^m \left[\mathcal{V} + (-1)^n \Theta_{_{\Phi}}^{\frac{1}{2}}\right]^{\frac{1}{2}}\hat{I}_{_{4}}$, again with the resulting related coefficient at Eq.~\eqref{ForthH2NN} canceling the denominator. In both limit cases, one does not have a pure state anymore, with $\rho_{mn} \equiv \hat{I}_{_{4}}$ reduced to a maximal statistical mixing.},
\begin{eqnarray}
\label{ForthH2NN}
\rho_{mn} &=& 
\frac{1}{4}
\left[\hat{I}_{_{{ 4}}} + \frac{\hat{H} -\upsilon \hat{I}_{_{{ 4}}}}{\varepsilon_{mn} -\upsilon}\right]
\left[\hat{I}_{_{{ 4}}} + \frac{\tilde{H}^2 -\mathcal{V} \hat{I}_{_{{ 4}}}}{(\varepsilon_{mn} -\upsilon)^2 - \mathcal{V}}\right]\nonumber\\
&=& 
\frac{1}{4}
\left[\hat{I}_{_{{ 4}}} + (-1)^m \frac{\tilde{H}}{\left[\mathcal{V} + (-1)^n \Theta_{_{\Phi}}^{\frac{1}{2}}\right]^{\frac{1}{2}}}\right]
\left[\hat{I}_{_{{ 4}}} + (-1)^n\frac{\mathcal{O}}{\Theta_{_{\Phi}}^{\frac{1}{2}}}\right].
\end{eqnarray}
Given that $[\rho_{mn},\,\hat{H}] = [\rho_{mn},\,\tilde{H}] = 0$, Eq.~\eqref{ForthH2NN} is a fine-tuned version of the {\em ansatz} 
\begin{eqnarray}
\label{ForthH2NNB}
\rho_{mn} &=& \sum_{i=0}^{\infty}\xi_i\,\tilde{H}^i,
\end{eqnarray}
which reduces to
\begin{eqnarray}
\label{ForthH2NBBB}
\rho_{mn} &=& \sum_{i=0}^{3}\xi_i\,\tilde{H}^i,
\end{eqnarray}
for the constraint Eqs.~\eqref{ThirdH} and \eqref{ThirdH3B}, through which powers of $\tilde{H}$, $\tilde{H}^i$ with $i>4$, are given in terms of some linear combination of lower powers, $\sum_{i=0}^{3}\zeta_i\,\tilde{H}^i$, which is also an evident straightforward consequence of the Cayley-Hamilton theorem.
Finding $\rho_{mn}$ consists in finding the coefficients $\{\xi_i\}$, for $i = 0,\,1,\,2,\,3$, typically constrained by $\Tr[\rho_{mn}]=1$, and by $\Tr[\rho_{mn}\rho_{pq}] = \delta_{mn}\delta_{pq}$ for orthonormalized pure eigenstates (see the Appendix II).

\subsubsection{Quantum Concurrence}

For bipartite states, $\rho_{mn}$, which are not simply written as \eqref{bipar}, some more elaborated criteria for quantifying entanglement are required.
For pure states in the four-dimensional composite Hilbert spaced identified by $\mathbb{H} = \mathbb{H}^{(1)} \otimes \mathbb{H}^{(2)}$ with $\mbox{dim}\, \mathbb{H}^{(1)} = \mbox{dim}\, \mathbb{H}^{(2)} = 2$, considering that reduced density matrices, identified by $\rho_{1,2} = \mbox{Tr}_{2,1}[\rho]$, have identical eigenvalues, if the state is entangled, then either $\rho_{1}$ or $\rho_{2}$ are mixed (the Schmidt theorem) \cite{Entanglement02}, and an entanglement entropy can be identified as the von Neumann entropy of the reduced state $\rho_{1,2} = \mbox{Tr}_{2,1}[\rho]$ \cite{Entanglement02, Entanglement03}, i.e.
\begin{equation}
E_{\mbox{\tiny vN}} [ \rho ] = S[\rho_2] = - \mbox{Tr}_2[\rho_2 \mbox{log}_2 \rho] = S[\rho_1] = - \mbox{Tr}_1[\rho_1 \mbox{log}_1 \rho].
\end{equation}
Often more conveniently applied, the quantum concurrence $\mathcal{C}[\rho]$, which quantifies the entanglement of formation \cite{Entanglement04}, is defined by
\begin{equation}\label{Novvva}
\mathcal{C}[\rho] = \mbox{max} \{\lambda_1 - \lambda_2 - \lambda_3 - \lambda_4, 0 \},
\end{equation}
 where $\lambda_1 > \lambda_2 > \lambda_3 > \lambda_4$ are the eigenvalues of the operator $$\sqrt{\sqrt{\rho}(\sigma^{(1)}_2 \otimes \sigma^{(2)}_2) \rho^*(\sigma^{(1)}_2 \otimes \sigma^{(2)}_2) \sqrt{\rho}}.$$ 

For pure states, as those from Eq.~\eqref{ForthH2NN}, when described according to the Fano decomposition \cite{Fano}, 
\begin{equation}
\rho = \frac{1}{4} \left[ \hat{I}_{_{{ 4}}} + A_{i}\,\hat{{\sigma}}_i \otimes \hat{I}_{_{{ 2}}} + B_{i}\,\hat{I}_{_{{ 2}}}\otimes \hat{{\sigma}}_j + \displaystyle \sum_{i,j = 1} ^3 W_{ij} \,\hat{\sigma}_i \otimes \hat{\sigma}_j \right].
\end{equation}
the quantum concurrence is evaluated by
\begin{equation}
\mathcal{C}[\rho] = \sqrt{1 - A ^2} = \sqrt{1 - B^2},
\end{equation}
with $A$ and $B$, the modulus of the Bloch vectors associated to each subsystem.

For the eigenstates from \eqref{ForthH2NN}, one has 
\begin{eqnarray}
\label{ForthH2NBBB}
A_i &=& (-1)^n\frac{\mathcal{A}_i}{\Theta_{_{\Phi}}^{\frac{1}{2}}}+(-1)^m\frac{\alpha_i}{\left[\mathcal{V} + (-1)^n \Theta_{_{\Phi}}^{\frac{1}{2}}\right]^{\frac{1}{2}}}+(-1)^{m+n}\frac{\mathcal{W}_{ik}\,\beta_k+\omega_{ik}\mathcal{B}_k}{\Theta_{_{\Phi}}^{\frac{1}{2}}\left[\mathcal{V} + (-1)^n \Theta_{_{\Phi}}^{\frac{1}{2}}\right]^{\frac{1}{2}}},\\
B_j &=& (-1)^n\frac{\mathcal{B}_j}{\Theta_{_{\Phi}}^{\frac{1}{2}}}+(-1)^m\frac{\beta_j}{\left[\mathcal{V} + (-1)^n \Theta_{_{\Phi}}^{\frac{1}{2}}\right]^{\frac{1}{2}}}+(-1)^{m+n}\frac{\alpha_k\,\mathcal{W}_{kj}+\mathcal{A}_k\omega_{kj}}{\Theta_{_{\Phi}}^{\frac{1}{2}}\left[\mathcal{V} + (-1)^n \Theta_{_{\Phi}}^{\frac{1}{2}}\right]^{\frac{1}{2}}},
\end{eqnarray}
With the above conditions, one can verify that $A ^2=B ^2$. Considering the simplifications yielded by Eqs.~\eqref{above4}-\eqref{ForthHSSBFim}, with $\mbox{\boldmath$\alpha$}\cdot\mbox{\boldmath$\omega$}\cdot\mbox{\boldmath$\beta$}=0$,
the quantum concurrence results in
\small\begin{equation}
\mathcal{C}[\rho_{mn}] =\left\{\frac{\Phi}{\Theta_{_{\Phi}}} - \frac{\mbox{\boldmath$\alpha$}^2\,\delta_{{\tiny{\mbox{\boldmath$\alpha$}\cdot\mbox{\boldmath$\omega$}}}}^{0} + \mbox{\boldmath$\beta$}^2\delta_{{\tiny{\mbox{\boldmath$\omega$}\cdot\mbox{\boldmath$\beta$}}}}^{0}}{\mathcal{V} + (-1)^n \Theta_{_{\Phi}}^{\frac{1}{2}}}\left[1+2(-1)^n\frac{\mbox{\boldmath$\beta$}^2\,\delta_{{\tiny{\mbox{\boldmath$\alpha$}\cdot\mbox{\boldmath$\omega$}}}}^{0} + \mbox{\boldmath$\alpha$}^2\delta_{{\tiny{\mbox{\boldmath$\omega$}\cdot\mbox{\boldmath$\beta$}}}}^{0}}{\mbox{\boldmath$\alpha$}^2\mbox{\boldmath$\beta$}^2\Theta_{_{\Phi}}^{\frac{1}{2}}}\left(\mbox{\boldmath$\alpha$}^2\mbox{\boldmath$\beta$}^2-\mbox{\boldmath$\alpha$}\cdot\mbox{\boldmath$\beta$}\det[\mbox{\boldmath$\omega$}_{_B}]\right) \right]^2\right\}^{\frac{1}{2}}.\end{equation}\normalsize

\subsubsection{TD ensembles}

Following the same procedure as for separate states, from Eqs.~\eqref{ForthH2BNN} and \eqref{ForthH2NN}, the thermal density constructed from the above obtained entangled states is identified by
\small\begin{eqnarray}
\varrho(T) &=& \frac{e^{-\frac{\upsilon}{k_{_B} T}}}{4\mathcal{Z}(T)}
\sum_{m,n=1,2}\exp\left[(-1)^{m+1} E_n/k_{_B} T\right]\left[\hat{I}_{_{{ 4}}} + (-1)^m \frac{\tilde{H}}{E_n}\right]
\left[\hat{I}_{_{{ 4}}} + (-1)^n\frac{\mathcal{O}}{\Theta_{_{\Phi}}^{\frac{1}{2}}}\right]
\nonumber\\
&=&
\frac{e^{-\frac{\upsilon}{k_{_B} T}}}{2\mathcal{Z}(T)}
\sum_{n=1,2}\left[
\cosh\left(E_n/k_{_B} T\right)\hat{I}_{_{{ 4}}} 
- \sinh\left(E_n/k_{_B} T\right) \frac{\tilde{H}}{E_n}\right\}\left[\hat{I}_{_{{ 4}}} + (-1)^n\frac{\mathcal{O}}{\Theta_{_{\Phi}}^{\frac{1}{2}}}\right]
\nonumber\\
&\equiv&
\frac{e^{-\frac{\upsilon}{k_{_B} T}}}{2\mathcal{Z}(T)}
\sum_{n=1,2}\exp\left(-\tilde{H}/k_{_B} T\right)\left[\hat{I}_{_{{ 4}}} + (-1)^n\frac{\mathcal{O}}{\Theta_{_{\Phi}}^{\frac{1}{2}}}\right]
\nonumber\\
&\equiv& \exp\left(-\hat{H}/k_{_B} T\right)/\mathcal{Z}(T),\qquad
\end{eqnarray}\normalsize
with $E_n=\left[\mathcal{V} + (-1)^n \Theta_{_{\Phi}}^{\frac{1}{2}}\right]^{\frac{1}{2}}$, which yields the partition function,
\begin{eqnarray}
\label{PF2}
\mathcal{Z}(T) &=& \Tr\left[\exp\left(-\hat{H}/k_{_B} T\right)\right]\nonumber\\
&=& \sum_{m,n=1,2}\exp\left[-\left(\upsilon+(-1)^{m+1} E_n\right)/k_{_B} T\right]
\nonumber\\
&=& 2\,\exp\left(-\upsilon/k_{_B} T\right)\sum_{n=1,2}\cosh\left(E_n/k_{_B} T\right)
\nonumber\\
&=& 2\,\exp\left(-\upsilon/k_{_B} T\right)
\left\{\cosh\left[\left(\mathcal{V} + \Theta_{_{\Phi}}^{\frac{1}{2}}\right)^{\frac{1}{2}}/k_{_B} T\right]+\cosh\left[\left(\mathcal{V} - \Theta_{_{\Phi}}^{\frac{1}{2}}\right)^{\frac{1}{2}}/k_{_B} T\right]\right\}.
\end{eqnarray}
Again, the partition function yields the associated quantum purity, written as 
\cite{Bernardini2020AA}
\begin{eqnarray}
\label{PF}
\mathcal{P}(T) &=&\frac{\cosh^2\left[\left(\mathcal{V} + \Theta_{_{\Phi}}^{\frac{1}{2}}\right)^{\frac{1}{2}}/k_{_B} T\right]+\cosh^2\left[\left(\mathcal{V} - \Theta_{_{\Phi}}^{\frac{1}{2}}\right)^{\frac{1}{2}}/k_{_B} T\right]-1}{\left\{\cosh\left[\left(\mathcal{V} + \Theta_{_{\Phi}}^{\frac{1}{2}}\right)^{\frac{1}{2}}/k_{_B} T\right]+\cosh\left[\left(\mathcal{V} - \Theta_{_{\Phi}}^{\frac{1}{2}}\right)^{\frac{1}{2}}/k_{_B} T\right]\right\}^2},
\end{eqnarray}
which is also consistent with the classical and pure state limits, $\mathcal{P}(\infty) = 1/4$ and $\mathcal{P}(0) = 1$, respectively.
Considering the hypothesis of having negative energy eigenstates, an ensemble of only positive energy eigenstates can be described by Eqs.~\eqref{PF2} and \eqref{PF} with $\cosh[(\dots)]$ replaced by $\exp[-(\dots)]/2$.

Furthermore, the quantum concurrence associated with $\varrho(T)$ can be straightforwardly obtained from its naive definition from Eq.~\eqref{Novvva}, since the operation
$$(\sigma^{(1)}_2 \otimes \sigma^{(2)}_2) \varrho(T)^*(\sigma^{(1)}_2 \otimes \sigma^{(2)}_2)$$
simply changes $\{\mbox{\boldmath$\alpha$},\,\mbox{\boldmath$\beta$}\}$ into $-\{\mbox{\boldmath$\alpha$},\,\mbox{\boldmath$\beta$}\}$ in $\tilde{H}$, and therefore,
$$ \sqrt{\sqrt{\varrho(T)}(\sigma^{(1)}_2 \otimes \sigma^{(2)}_2) \varrho(T)^*(\sigma^{(1)}_2 \otimes \sigma^{(2)}_2) \sqrt{\varrho(T)}}= \frac{e^{-\frac{\upsilon}{k_{_B} T}}}{\mathcal{Z}(T)}\exp\left[\frac{\omega_{ij}\,\hat{\sigma}_i \otimes \hat{\sigma}_j}{k_{_B} T}\right].$$
Given that the eigenvalues of $\omega_{ij}\,\hat{\sigma}_i \otimes \hat{\sigma}_j$ are given by $\pm\sqrt{\Tr[\mbox{\boldmath$\omega$}_B^2]\pm 2\vert\det[\mbox{\boldmath$\omega$}_B]\vert}$,
one has\\ 
\small\begin{equation}\label{concm}
\mathcal{C}[\varrho(T)] = \frac{\mbox{max} \left\{
\sinh\left[\sqrt{\Tr[\mbox{\boldmath$\omega$}_B^2]+2\vert\det[\mbox{\boldmath$\omega$}_B]\vert}/k_{_B} T\right]
-\cosh\left[\sqrt{\Tr[\mbox{\boldmath$\omega$}_B^2]-2\vert\det[\mbox{\boldmath$\omega$}_B]\vert}/k_{_B} T\right],\,0
\right\}}{\left\{\cosh\left[\left(\mathcal{V} + \Theta_{_{\Phi}}^{\frac{1}{2}}\right)^{\frac{1}{2}}/k_{_B} T\right]+\cosh\left[\left(\mathcal{V} - \Theta_{_{\Phi}}^{\frac{1}{2}}\right)^{\frac{1}{2}}/k_{_B} T\right]\right\}
},
\end{equation}\normalsize
which corresponds to a general form for the quantum concurrence related to $SU(2)\otimes SU(2)$ TD ensembles subjected to the constraints discussed here.

\section{The Bernal stacked graphene}

As a subtle platform for testing the above discussed framework, the single particle excitations of bilayer graphene structures can be entirely described by a $SU(2) \otimes SU(2)$ Hamiltonian.
Considering the bilayer graphene in its most stable configuration, the AB (or Bernal) stacking \cite{graph03, graph04}, one can assume that the $SU(2) \otimes SU(2)$ quantum correlational content generated by Hamiltonian interactions can be re-interpreted in terms of {\em lattice-layer} associated parameters.

As previously noticed, the inclusion of Hamiltonian interaction contributions into the $SU(2) \otimes SU(2)$ dynamics affects the correlational content of the bi-spinor states associated to each $SU(2)$ Hilbert space.

For the so-called AB tight binding Hamiltonian used to describe the bilayer graphene, the geometry of the AB stacking (or Bernal stacking) corresponds to two layers of graphene arranged such that half of the atoms of the upper layer are localized above half of the atoms of the lower layer (dimer sites), while the other half atoms are localized above the center of the lower honeycombs (non-dimer sites) \cite{graph03,graph04,PRBPRB18}.

The tight binding Hamiltonian is given by \cite{graph03,graph04,OldGraph01,OldGraph02,PRBPRB18}
\begin{eqnarray}
\label{tightbinding}
\hat{\mathcal{H}}_{AB} = &-& t \displaystyle \sum_{\bm{k}} \left[ \, \Gamma(\bm{k}) \hat{a}_{1\bm{k}} ^\dagger \hat{b}_{1\bm{k}} + \Gamma(\bm{k}) \hat{a}_{2 \bm{k}}^\dagger \hat{b}_{2 \bm{k}} + \mbox{h.c.} \,\right] \nonumber \\
 &+& t_{\bot} \displaystyle \sum_{\bm{k}} \left[ \, \hat{b}_{1 \bm{k}}^\dagger \hat{a}_{2 \bm{k}} + \hat{a}_{2 \bm{k}}^\dagger \hat{b}_{1 \bm{k}} \, \right] - t_3 \displaystyle \sum_{\bm{k}} \left[\, \Gamma(\bm{k}) \hat{b}_{2 \bm{k}}^\dagger \hat{a}_{1 \bm{k}} + \Gamma^* (\bm{k}) \hat{a}_{1 \bm{k}}^\dagger \hat{b}_{2 \bm{k}} \, \right] \nonumber \\
 &+&t_4 \displaystyle \sum_{\bm{k}} \left[ \, \Gamma(\bm{k})(\hat{a}_{1 \bm{k}}^\dagger \hat{a}_{2 \bm{k}} + \hat{b}_{1 \bm{k}}^\dagger \hat{b}_{2 \bm{k}}) + \mbox{h.c.} \, \right],
\end{eqnarray}
where, according to the geometry, $t$ corresponds to the hopping between next-neighbors in the same layer; $t_3$ ($t_\bot$) describes the hopping from a (non-) dimer site to its nearest (non-) dimer site; and $t_4$ is the hopping from a dimer to the nearest non-dimer site. One also identifies $\hat{a}^\dagger(\hat{b}^\dagger)_{i \bm{k}}$ as the creation operator for an excitation on the $a({b})$ lattice in the $i$ layer with wave vector $\bm{k}$, and $\Gamma(\bm{k}) = \displaystyle \sum_{j=1} ^3 e^{ i \bm{k} \cdot \bm{\delta}_{j}}$ is given in terms of the next-neighbor vectors, $\delta_i$,
\begin{equation}
\label{gamma}
\Gamma(\bm{k}) = 2\exp[-i\,k_x\,\lambda/2]\,\cos[\sqrt{3} k_y\,\lambda/2] + \exp[-i\,k_x\,\lambda],
\end{equation}
where $$\bm{\delta}_{1,2} = \left( -\lambda/2, \, \pm \lambda \sqrt{3}/2 \right), \hspace{0.5 cm} \bm{\delta}_3 = \left(\lambda, 0 \right),$$ 
and $\lambda$ is the lattice parameter \cite{graph03,graph04,OldGraph01,OldGraph02,PRBPRB18}.

By setting $t_4=0$ ($\upsilon = 0$ for diagonal contributions), in the $\bm{k}$ space, the Hamiltonian $\hat{\mathcal{H}}_{AB}$ is then written in the basis $\{ \vert A1 (\bm{k}) \rangle, \vert B1 (\bm{k}) \rangle, \vert A2 (\bm{k}) \rangle, \vert B2 (\bm{k}) \rangle \}$ ($\vert \alpha _i (\bm{k}) \rangle = \hat{\alpha}_{i \bm{k}} ^\dagger \vert 0 \rangle$), where $A1(2)$ and $B1(2)$ identify the sub-lattices of layer 1(2). This leads to
\begin{equation}
\label{ABHamiltonian}
\hat{\mathcal{H}}_{AB} =\left[ \, \begin{array}{cccc}
0 & - t \Gamma(\bm{k}) & 0 & - t_3 \Gamma^*(\bm{k})\\
- t \Gamma^*(\bm{k}) & 0 & t_\bot & 0\\
0 & t_\bot & 0& -t\Gamma (\bm{k})\\
- t_3 \Gamma(\bm{k}) & 0 & - t \Gamma^*(\bm{k}) & 0
\end{array} \right].
\end{equation}
Assuming additional interactions associated to an energy gapping \cite{graph03, Predin}, a mass term, $\hat{\mathcal{H}}_m$, and a bias voltage term, $\hat{\mathcal{H}}_{\Lambda}$, are respectively given by
\begin{equation}
\label{massHamiltonian}
\hat{\mathcal{H}}_{m}= \mbox{diag}\{ m, \, -m, \, m, \, -m\},
\end{equation}
\begin{equation}
\label{biasvoltageHamiltonian}
\hat{\mathcal{H}}_{\Lambda}= \mbox{diag} \left\{ \frac{\Lambda}{2}, \, \frac{\Lambda}{2}, \, - \frac{\Lambda}{2}, \, - \frac{\Lambda}{2} \right\}.
\end{equation}
Considering Hamiltonian interactions involving both mass and bias voltage, the total Hamiltonian can be written as
\begin{eqnarray}\label{E04T}
\hat{\mathcal{H}} &=& \hat{\mathcal{H}}_{AB} + \hat{\mathcal{H}}_m + \hat{\mathcal{H}}_{\Lambda}\nonumber\\ 
 &=& \alpha_3 \,\hat{\sigma}_3^{(1)} \otimes \hat{I}^{(2)} + \beta_j \, \hat{I}^{(1)} \otimes {\hat{\sigma}}_j^{(2)} + \omega_{1j} \, \hat{\sigma}_1 ^{(1)} \otimes {\hat{\sigma}}_j^{(2)} + \omega_{2j} \, \hat{\sigma}_2^{(1)} \otimes {\hat{\sigma}}_j^{(2)},
\end{eqnarray}
from which one identifies the following one-to-one correspondence between graphene and $SU(2) \otimes SU(2)$ parameters,
\begin{eqnarray}
\label{relations}
\alpha_3 &\leftrightarrow& \frac{\Lambda}{2};
\nonumber\\
\{\beta_1,\,\beta_2,\,\beta_3\} &\leftrightarrow& \{- t \,\mbox{Re}[\Gamma (\bm{k})],\, t \,\mbox{Im}[\Gamma(\bm{k})],\, m\};
\nonumber\\
\{\omega_{11},\,\omega_{12},\,\omega_{13}\} &\leftrightarrow& \Big\{\frac{t_\bot - t_3 \,\mbox{Re}[\Gamma (\bm{k})]}{2},\, - \frac{t_3 \,\mbox{Im}[\Gamma(\bm{k})]}{2} ,\,0\Big\};
\nonumber\\
\{\omega_{21},\,\omega_{22},\,\omega_{23}\} &\leftrightarrow& \Big\{-\frac{ t_3 \,\mbox{Im}[\Gamma(\bm{k})]}{2},\, \frac{t_\bot + t_3 \,\mbox{Re}[\Gamma(\bm{k})]}{2},\,0\Big\};
\end{eqnarray}
from which it is straightforward to notice that constraint conditions over $\mbox{\boldmath$\alpha$}\cdot\mbox{\boldmath$\omega$}\cdot\mbox{\boldmath$\beta$}$ is satisfied. Since $\bm{\alpha}\cdot\bm{\omega}=0$, the associated Hamiltonian eigenstates, $\rho_{mn}$, with $m,\,n = 1,\,2$, can be identified. Their quantum correlational content, as well as their thermodynamic ensemble properties, as discussed in the previous section, can also be straightforwardly obtained.
From \eqref{relations}, one has $\bm{\alpha}^2 = \Lambda^2/4$, $\bm{\beta}^2 = t^2\,\vert\Gamma (\bm{k})\vert^2 + m^2$, and $\Tr[\mbox{\boldmath$\omega$}\cdot\mbox{\boldmath$\omega$}^T] = (t_\bot^2+t_3^2\,\vert\Gamma (\bm{k})\vert^2)/2$, which, once inserted into Eq.~\eqref{ThirdH2}, gives
\begin{equation}
\label{ThirdH2Fim}
\mathcal{V} = m^2 + \frac{\Lambda^2}{4} + \frac{1}{2}(2t^2+t_\bot^2+t_3^2\,\vert\Gamma (\bm{k})\vert^2).
\end{equation}

In addition, one has $\det[\mbox{\boldmath$\omega$}_{_{B}}]=(t_\bot^2-t_3^2\,\vert\Gamma (\bm{k})\vert^2)/4$, from which 
\begin{equation}
\label{ThirdH2Fim}
\Phi = \frac{1}{4}\left(2 m\Lambda - t_\bot^2+t_3^2\,\vert\Gamma (\bm{k})\vert^2\right)^2 + \Lambda^2\,t^2 \left(\vert\Gamma (\bm{k})\vert^2 + 2 \mbox{Re}[\Gamma (\bm{k})]\mbox{Im}[\Gamma (\bm{k})] \right).\end{equation}
Finally, with 
\begin{eqnarray}
\label{ThirdH2Fim3}
\mbox{\boldmath$\beta$}\cdot\mbox{\boldmath$\omega$}\cdot\mbox{\boldmath$\omega$}^T\hspace{-.15cm}\cdot\mbox{\boldmath$\beta$}&=&
\frac{t^2}{4}\left\{\vert\Gamma (\bm{k})\vert^2(t_\bot^2+t_3^2\,\vert\Gamma (\bm{k})\vert^2)-2t_\bot t_3 \left[\vert\Gamma (\bm{k})\vert^2-4(\mbox{Im}[\Gamma (\bm{k})])^2\right]\,\mbox{Re}[\Gamma (\bm{k})]\right\},\qquad
\end{eqnarray}
$\Theta_{_{\Phi}}$ from Eq.~\eqref{ForthHSSBFim}, and $\Gamma (\bm{k})$ from \eqref{gamma}, all quantum observables discussed in the previous section can be straightforwardly computed.

\begin{figure}
\includegraphics[width= 16 cm]{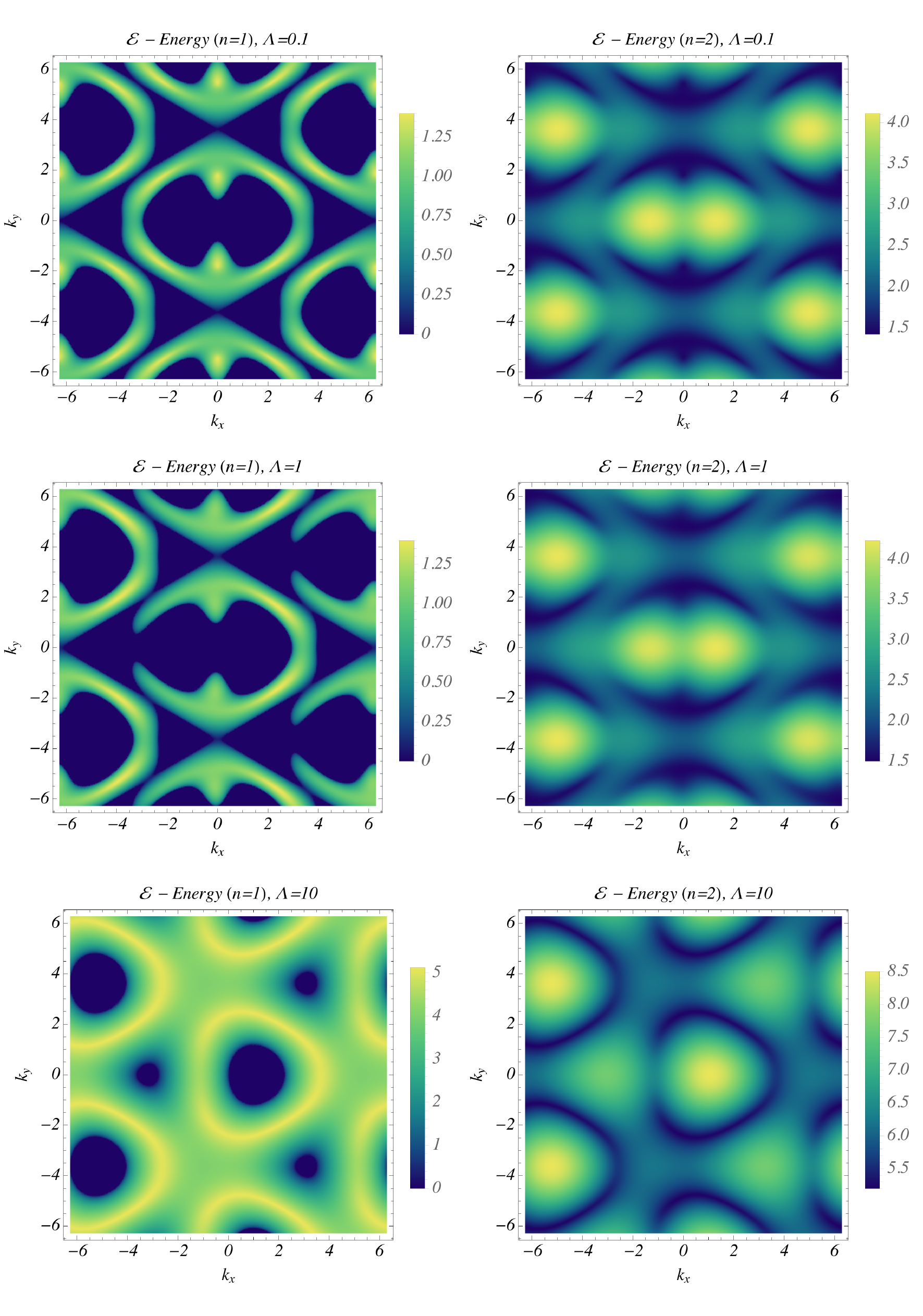}
\renewcommand{\baselinestretch}{1.0}
\vspace{-.5 cm}\caption{Positive energy bands for the branches $n=1$ (first column) and $n=2$ (second column) for $\Lambda = 0.1$ (first row), $1$ (second row), and $10$ (third row).}
\label{concurrencecomp1}
\end{figure}

\begin{figure}
\includegraphics[width= 16 cm]{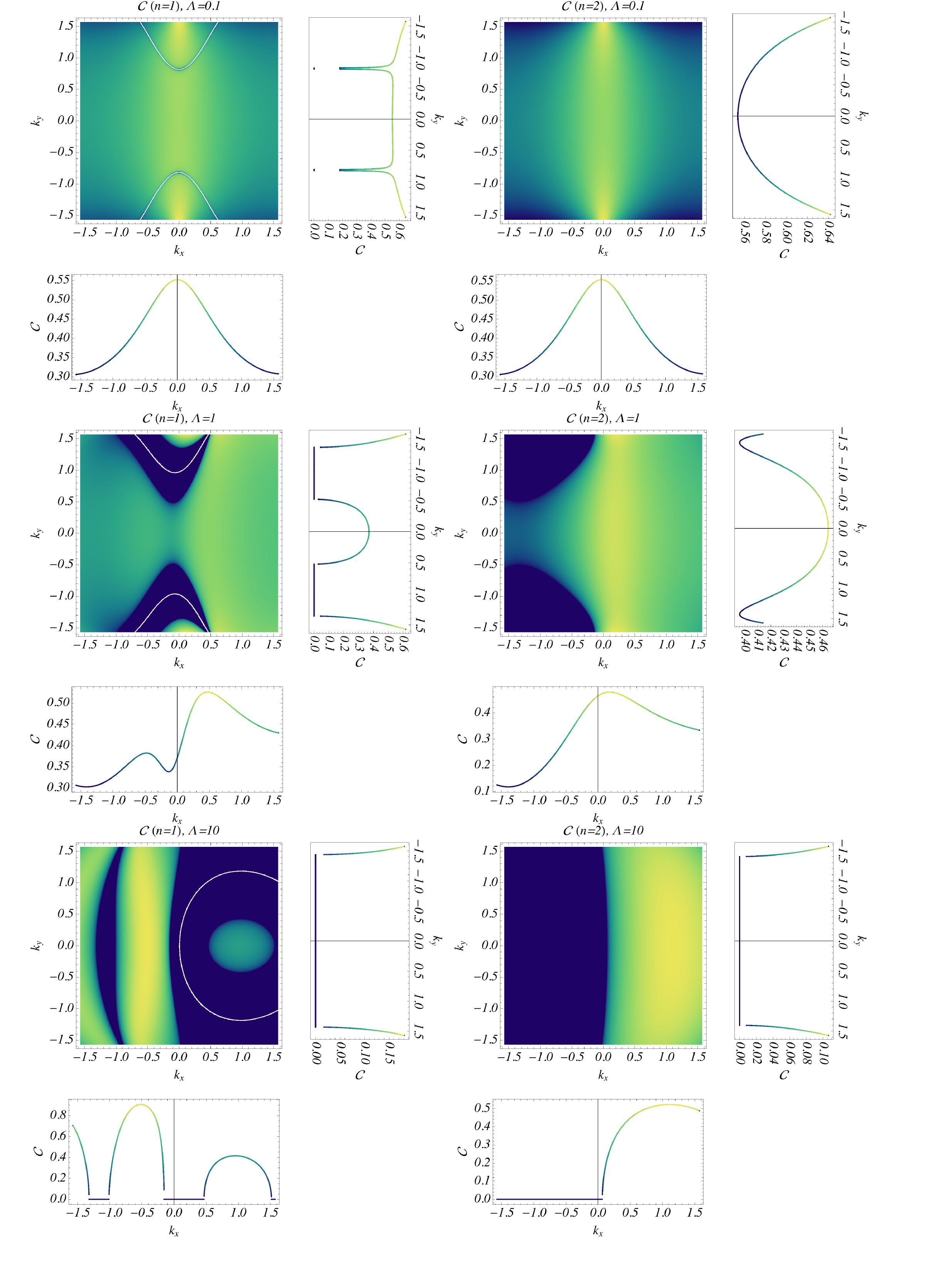}
\renewcommand{\baselinestretch}{1.0}
\caption{Quantum concurrence in the first Brillouin zone for the branches $n=1$ (first column) and $n=2$ (second column) for $\Lambda = 0.1$ (first row), $1$ (second row), and $10$ (third row), in terms of a {\em blue-green-yellow} scheme quantified according to the sectional plots.}
\label{concurrencecomp2}
\end{figure}

Qualitatively, a subtle driver for introducing contrasting phenomenology is the bias voltage, $\Lambda$.
For this reason, the following results are presented in terms of $\Lambda$ and $\mathbf{k}$ parameters, for $m=0$, $t=t_\bot=t_3=1$. Fig.~\ref{concurrencecomp1} depicts the positive energy bands for the branches $n=1$ (first column) and $n=2$ (second column) for $\Lambda = 0.1$ (first row), $1$ (second row), and $10$ (third row). Following a {\em blue-green-yellow} scale, for the branch $n=1$, the dark blue halos denote the Dirac point spreading at the zero point energy, denoting the conducing behavior (positive and negative energy bands do encompass each other). For the branch $n=2$, the dark blue region denote the minimal positive energy (for which there is a corresponding symmetric negative one), denoting the non-conducing behavior (positive and negative energy bands do not intercept each other). Correspondently, Fig.~\ref{concurrencecomp2} depicts the quantum concurrence for the same set of parameters. In this case, blue halos and white tiny regions denote separable state configurations ($\mathcal{C} = 0$). White regions correspond to non-real eigenvalues at Eq.~\eqref{Novvva} (separable states, with $\mathcal{C} = 0$) though they follow an analytic continuation to the values for $k_x$ and $k_y$ (cf. depicted by Fig.~\ref{concurrencecomp1}).

Just to end up, for the associated TD ensemble, the quantum concurrence Eq.~\eqref{concm} is given by
\begin{equation}
\mathcal{C}[\varrho(T)] = \frac{\mbox{max} \left\{
\sinh\left[t_\bot/k_{_B} T\right]
-\cosh\left[t_3\,\vert\Gamma (\bm{k})\vert/k_{_B} T\right],\,0
\right\}}{\left\{\cosh\left[\left(\mathcal{V} + \Theta_{_{\Phi}}^{\frac{1}{2}}\right)^{\frac{1}{2}}/k_{_B} T\right]+\cosh\left[\left(\mathcal{V} - \Theta_{_{\Phi}}^{\frac{1}{2}}\right)^{\frac{1}{2}}/k_{_B} T\right]\right\}},
\end{equation}
from which, given the number of free parameters, a deeper phenomenological analysis, as that performed in Ref.~\cite{MeuPRB}, can be clearly persecuted.

Of course, the point here is the manipulability provided by the general form of $SU(2) \otimes SU(2)$ associated quantum observables, which simplify such analysis and can be straightforwardly extended to the systematic computation of physical observables related to several mesoscopic systems, which include: photonic graphene, magnetically trapped ions, Weyl- and Dirac semi-metals, paramagnetic honeycomb ruthenates, and generic Dirac phononic and magnonic systems \cite{Kumar20}.

\section{Conclusions}

Generic $SU(2)\otimes SU(2)$ Hamiltonian eigensystems were solved through systematic manipulations of quartic polynomial equations for which algebraic solutions could be provided by means of the Cardano-Ferrari's method \cite{Cardano}. 
{\em Ans\"atze} for constructing separable and entangled eigenstate basis, depending on the quartic equation coefficients, were evaluated. Besides a generic algorithm for describing properties of separable and entangled states, their associated thermodynamic statistical ensemble properties were identified. 
Results were specialized to a $SU(2)\otimes SU(2)$ structure emulated by lattice-layer degrees of freedom of the Bernal stacked graphene, in a context that can be extended to several mesoscopic scale systems for which the onset from $SU(2) \otimes SU(2)$ Hamiltonians has been assumed.
In fact, the manipulability provided by the general form of $SU(2) \otimes SU(2)$ associated quantum observables can be straightforwardly extended to the systematic computation of physical observables related to several mesoscopic systems, which include: photonic graphene, magnetically trapped ions, Weyl- and Dirac semi-metals, paramagnetic honeycomb ruthenates, and generic Dirac phononic and magnonic systems \cite{Kumar20}, all encompassed by our next investigations.

{\em Acknowledgments -- The work of AEB is supported by the Brazilian Agencies FAPESP (Grant No. 2023/00392-8, S\~ao Paulo Research Foundation (FAPESP))) and CNPq (Grant No. 301485/2022-4). RdR thanks to The S\~ao Paulo Research Foundation -- FAPESP (Grants No. 2021/01089-1 and No. 2022/01734-7), and to the National Council for Scientific and Technological Development -- CNPq (Grants No. 303742/2023-2 and No. 401567/2023-0), for partial financial support.}

\section*{Appendix I -- Group representation of elementary $SU(2)\otimes SU(2)$ structures}

In particular, considering that $SU(2)\otimes SU(2)$ are described by a subset of $SL(2,\mathbb{C})\otimes SL(2,\mathbb{C})$, at least two {\em inequivalent} subsets of $SU(2)$ generators are identified with $SU(2)\otimes SU(2) \subset SL(2,\mathbb{C})\otimes SL(2,\mathbb{C})$, in which each generator has its own irreducible representation, $irrep (su_{\xi}(2)\oplus su_{\chi}(2))$. For instance, a {\em spinor} $\xi$ described by $(\frac{1}{2},\,0)$ transforms as a {\em doublet} - object of the fundamental representation - of $SU_{\xi}(2)$, and as a singlet - object ``transparent'' to transformations - of the $SU_{\chi} (2)$ group. Under the notation $(\mbf{dim}(SU_{\xi}(2)),\mbf{dim}(SU_{\chi}(2)))$, the {\em spinor} $\xi$ is an object described by the algebra representation (\mbf{2},\,\mbf{1}). Conversely, by the same reason, the {\em spinor} $\chi$, $(0,\,\frac{1}{2})$, which transforms as a {\em singlet} of $SU_{\xi}(2)$ and as a {\em doublet} of $SU_{\chi}(2)$, is represented (\mbf{1},\,\mbf{2}).
Following the same procedure, additional representations can be identified according to the Poincar\'e classes they belong:
$(\mbf{1},\mbf{1})$ -- a {\em scalar} or {\em singlet}, with angular momentum projection $j = 0$;
$(\mbf{2},\mbf{1})$ -- a {\em spinor} $(\frac{1}{2},\,0)$, usually quoted as {\em left-handed}, with angular momentum projection $j = 1/2$;
$(\mbf{1},\mbf{2})$ -- a {\em spinor} $(0,\,\frac{1}{2})$, usually quoted as {\em right-handed}, with angular momentum projection $j = 1/2$;
$(\mbf{2},\mbf{2})$ -- a {\em vector} or {\em doublet}, with angular momentum projection $j = 0$ and $j = 1$; where just those more elementary have been considered. Using these representations, all spinors in Minkowski spacetime can be classified into six sets, according to the Lounesto's spinor classification, with a wide range of new physical aspects and applications in field theory \cite{daRocha:2011yr,Cavalcanti:2014uta,Bonora:2017oyb,Lopes:2018cvu}, paving prime candidates for describing dark matter \cite{deGracia:2023yit,DaRocha:2020oju}. 

Assuming that such ground objects of an {\em irrep} can be used to construct more complex structures also classified by the Poincar\'e group structure\footnote{For instance, with respect to the representations of $SL(2,\mathbb{C})$, one has
$ (\mbf{1},\mbf{2}) \otimes (\mbf{1},\mbf{2}) \equiv (\mbf{1},\mbf{1}) \oplus (\mbf{1},\mbf{3})$,
a that composes Lorentz tensors like
\begin{equation}
C_{\alpha\beta}\bb{\mt{x}} = \epsilon_{\alpha\beta} D\bb{\mt{x}} + G_{\alpha\beta}\bb{\mt{x}},
\end{equation}
where $D\bb{\mt{x}}$ is a scalar, and $G_{\alpha\beta} = G_{\beta\alpha}$ is a totally symmetric tensor, or even
$ (\mbf{2},\mbf{1}) \otimes (\mbf{1},\mbf{2}) \equiv (\mbf{2},\mbf{2})$,
such that
$ (\mbf{2},\mbf{2}) \otimes (\mbf{2},\mbf{2}) \equiv (\mbf{1},\mbf{1}) \oplus (\mbf{1},\mbf{3}) \oplus (\mbf{3},\mbf{1}) \oplus (\mbf{3},\mbf{3})
$,
that are read as Lorentz tensors like
\begin{equation}
\varphi^{\mu\nu}\bb{\mt{x}} = A^{\mu\nu}\bb{\mt{x}} + S^{\mu\nu}\bb{\mt{x}} + \frac{1}{4}g^{\mu\nu} \Theta\bb{\mt{x}},
\end{equation}
which correspond to a decomposition into smaller {\em irreps}, with: $A^{\mu\nu} \equiv (\mbf{1},\mbf{3}) \oplus (\mbf{3},\mbf{1})$, a totally anti-symmetric tensor by the index interchange $\mu\leftrightarrow \nu$; $S^{\mu\nu}\equiv (\mbf{3},\mbf{3})$, a totally symmetric one by the index interchange $\mu\leftrightarrow \nu$; and $\Theta \equiv (\mbf{1},\mbf{1})$ as a Lorentz scalar multiplying the metric tensor, $g^{\mu\nu}$.},
the Hamiltonian dynamics described through a group representation described by a direct product between two algebras which compose a subset of the group $SL(2,\mathbb{C})\otimes SL(2,\mathbb{C})$, the group $SU(2)\otimes SU(2)$, emerge as a subtle driver of quantum correlations.

\section*{Appendix II -- Orthonormalization and pure eigenstate verification -- Entangled solutions}

From Eq.~\eqref{ForthH2NN} one has
\small\begin{eqnarray}
\label{ForthH2NNF}
\rho_{mn}\rho_{pq} &=& 
\frac{1}{16}
\left[\hat{I}_{_{{ 4}}} + (-1)^m \frac{\tilde{H}}{\left[\mathcal{V} + (-1)^n \Theta_{_{\Phi}}^{\frac{1}{2}}\right]^{\frac{1}{2}}}\right]
\left[\hat{I}_{_{{ 4}}} + (-1)^p \frac{\tilde{H}}{\left[\mathcal{V} + (-1)^q \Theta_{_{\Phi}}^{\frac{1}{2}}\right]^{\frac{1}{2}}}\right]
\nonumber\\
&&\quad \times \left[\hat{I}_{_{{ 4}}} + (-1)^n\frac{\mathcal{O}}{\Theta_{_{\Phi}}^{\frac{1}{2}}}\right]
\left[\hat{I}_{_{{ 4}}} + (-1)^q\frac{\mathcal{O}}{\Theta_{_{\Phi}}^{\frac{1}{2}}}\right]\nonumber
\end{eqnarray}\normalsize
\small\begin{eqnarray}
\label{ForthH2NNF}
&=& 
\frac{1}{16}
\left[\hat{I}_{_{{ 4}}} + (\dots)\tilde{H}+ (-1)^{m+p} \frac{\tilde{H}^2}{\left[\mathcal{V}^2 + (-1)^{n+q} \Theta_{_{\Phi}} +(-1)^{n}\,2\delta_{nq}\mathcal{V}\Theta_{_{\Phi}}^{\frac{1}{2}}\right]^{\frac{1}{2}}}\right]
\nonumber\\
&&\quad \times \left[\hat{I}_{_{{ 4}}} +(-1)^n\,2\delta_{nq}\frac{\mathcal{O}}{\Theta_{_{\Phi}}^{\frac{1}{2}}} + (-1)^{n+q}\frac{\mathcal{O}^2}{\Theta_{_{\Phi}}}\right]
\nonumber\\
&=& 
\frac{1}{16}
\left[\hat{I}_{_{{ 4}}} + (\dots)\tilde{H}+ (-1)^{m+p} \frac{\tilde{H}^2}{\left[\mathcal{V}^2 + (-1)^{n+q} \Theta_{_{\Phi}} +(-1)^{n}\,2\delta_{nq}\mathcal{V}\Theta_{_{\Phi}}^{\frac{1}{2}}\right]^{\frac{1}{2}}}\right]
\nonumber\\
&&\quad \times\left[(1+(-1)^{n+q})\hat{I}_{_{{ 4}}} +(-1)^n\,2\delta_{nq}\frac{\mathcal{O}}{\Theta_{_{\Phi}}^{\frac{1}{2}}}\right]
\nonumber\\
&=& 
\frac{\delta_{nq}}{8}
\left[\hat{I}_{_{{ 4}}} + (\dots)\tilde{H}+ (-1)^{m+p} \frac{\tilde{H}^2}{\mathcal{V} + (-1)^n \Theta_{_{\Phi}}^{\frac{1}{2}}}\right]\left[\hat{I}_{_{{ 4}}} + (-1)^n\frac{\mathcal{O}}{\Theta_{_{\Phi}}^{\frac{1}{2}}}\right],
\end{eqnarray}\normalsize
for $m,\,n,\,p,\,q = 1,\,2$.
Since $\Tr[\tilde{H}] = \Tr[\tilde{H}\mathcal{O}] = \Tr[\mathcal{O}] =0$, $\Tr[\tilde{H}^2] = 4 \mathcal{V}$ and $\Tr[\mathcal{O}^2] = 4 \Theta_{_{\Phi}}$, one has
\small\begin{eqnarray}
\label{ForthH2NNB}
\Tr[\rho_{mn}\rho_{pq}] &=& 
\frac{\delta_{nq}}{8}
\left[\Tr[\hat{I}_{_{{ 4}}}] + (-1)^{m+p} \frac{\Tr[\tilde{H}^2]+ (-1)^n \Theta_{_{\Phi}}^{-\frac{1}{2}}\Tr[\tilde{H}^2\mathcal{O} ]}{\mathcal{V} + (-1)^n \Theta_{_{\Phi}}^{\frac{1}{2}}}\right]\nonumber\\
&=& 
\frac{\delta_{nq}}{8}
\left[4 + (-1)^{m+p} \frac{4\mathcal{V}+ (-1)^n \Theta_{_{\Phi}}^{-\frac{1}{2}}\Tr[\mathcal{O}^2+\mathcal{O}\mathcal{V}]}{\mathcal{V} + (-1)^n \Theta_{_{\Phi}}^{\frac{1}{2}}}\right]
\nonumber\\
&=& 
\frac{\delta_{nq}}{2}
\left[1 + (-1)^{m+p} \frac{\mathcal{V} + (-1)^n \Theta_{_{\Phi}}^{\frac{1}{2}}}{\mathcal{V} + (-1)^n \Theta_{_{\Phi}}^{\frac{1}{2}}}\right]
\nonumber\\
&=& 
\delta_{mp}\delta_{nq},\end{eqnarray}\normalsize
which implies into an orthonormalized basis of pure eigenstates, $\{\rho_{11},\,\rho_{12},\,\rho_{21},\,\rho_{22}\}$.

\end{document}